\documentclass{aa}
\usepackage{epsfig}

\newcommand{\ud}[0]{\mathrm{d}}

\begin{document}
\def\h50{h$_{50}^{-1}${}}
\def\kms{km~s$^{-1}${}}

\title{A study of dark matter halos and gas properties in clusters of
galaxies from ROSAT data.  \thanks{Based on data obtained from the
public archive of ROSAT at the Max-Planck-Institut f\"ur
Extraterrestrische Physik.}}

\author{
  R.~Demarco \inst{1,2}
\and
  F.~Magnard \inst{2}
\and
  F.~Durret \inst{2}
\and
  I. M\'arquez  \inst{3}
}
\offprints{R. Demarco (\sl{demarco@iap.fr}) }
\institute{
        ESO-European Southern Observatory, Karl-Schwarzschild Str. 2,
 D-85748 Garching b. M\"unchen, Germany
\and 
	Institut d'Astrophysique de Paris, CNRS, 98bis Bd Arago, 
F-75014 Paris, France 
\and
        Instituto de Astrof\'\i sica de Andaluc\'\i a (C.S.I.C.), 
Apartado 3004 , E-18080 Granada, Spain
}

\date{Received February 14, 2003}

\authorrunning{R. Demarco et al.}  

\titlerunning{Dark matter halos and gas properties in clusters}

\abstract{ Self-gravitating systems such as elliptical galaxies appear
to have a constant integrated specific entropy and obey a scaling law
relating their potential energy to their mass.  These properties can
be interpreted as due to the physical processes involved in the
formation and evolution of these structures. Dark matter halos
obtained through numerical simulations have also been found to obey a
scaling law relating their potential energy to their mass with the
same slope as for ellipticals, and very close to the expected value
predicted by theory. Since the X-ray gas in clusters is weakly
dissipative, we test here the hypothesis that it verifies similar
properties. Comparable properties for the dark matter component are
also investigated.\\ With this aim, we have analyzed ROSAT-PSPC images
of 24 clusters, and fit a S\'ersic law to their X--ray surface
brightness profiles. We found that: 1)~the S\'ersic law parameters
(intensity, shape and scale) describing the X-ray gas emission are
correlated two by two, with a strong correlation between the shape and
scale parameters; 2)~the hot gas in all these clusters roughly has the
same integrated specific entropy, although a second order correlation
between this integrated specific entropy and both the gas mass
and the dynamical mass is observed; 3) a scaling law links the cluster
potential energy to its total mass, with the same slope as that
derived for elliptical galaxies and for dark matter halo
simulations. Comparable relations are obtained for the dark matter
component. All these correlations are probably the consequence of
the formation and evolution processes undergone by clusters of
galaxies.   }

\maketitle

\section{Introduction}

Clusters of galaxies are known to be the largest gravitationally bound
objects in the Universe. The amplification of primordial density
fluctuations by gravity is thought to be the origin of structure
formation, however the details of the formation process are not yet
well understood and the study of the structure and properties of dark
matter halos and of the intra cluster plasma in virialized systems can
give important clues to understand the physics involved in the
formation and evolution of galaxy clusters. Nowadays, these studies
have undergone great improvements with the developement of advanced
observational facilities and techniques, together with the progress of
numerical simulations.

Many works have been developed during the last decades on this
respect. Secondary infall and the effects of this process on the
cluster structure were discussed by Gunn \& Gott (1972), and
self-similar solutions for dark matter halos and gas were studied in
numerical simulations carried out e.g. by Fillmore \& Goldreich
(1984), Bertschinger (1985), Teyssier et al. (1997) and Subramanian
(2000). Cold Dark Matter (CDM) studies based on high-resolution N-body
simulations performed by Navarro, Frenk \& White (1996, 1997) suggest
a cuspy and universal dark matter (DM) density profile in galaxies and
clusters of galaxies, independent of mass scale and cosmology; this
result is contradicted by Jing \& Suto (2000). However, some important
observational facts seem not to be reproduced by these studies:
numerical simulations based on the CDM scenario predict density
profiles with steep inner slopes which fail to reproduce the rotation
curves of low surface brightness (LSB) galaxies (Flores \& Primack
1994; Moore et al. 1999).  Although other works claim that cuspy DM
profiles are consistent with the available data for dwarfs and LSB
galaxies (van den Bosch \& Swaters 2001), microlensing studies towards
the center of our galaxy also support the incompatibility between CDM
simulations and observational evidence (Binney \& Evans 2001). To
explain the discrepancy concerning the central slopes of DM halos
(Navarro, Frenk \& White 1997; Moore et al. 1999), the lack of
sufficient resolution in the central regions of simulated halos has
been proposed; Point Spread Function (PSF) effects together with
insufficient resolution on galaxy rotation curves (van den Bosch \&
Swaters 2001) may be also at the origin of the discrepancy between
models and observations.

Observations in the X-ray band provide valuable information on
the hot Intra Cluster Medium (ICM). A popular model
used to fit the spatial distribution of the X-ray gas is the so called
$\beta$-model (Cavaliere \& Fusco-Femiano 1976; Sarazin 1988) which
assumes an isothermal ICM. However, this model may not be good enough
to describe the cluster gas component, since the isothermality of the
ICM is still rather controversial. Cooling flows are known to produce
a drop of the gas temperature towards the center of the cluster;
besides, outside the cooling flow region things are still not clear.
Temperature profiles based on ASCA (White 2000) and ROSAT (Irwin,
Bregman \& Evrard 1999) data were found to be consistent with the
isothermal hypothesis. On the other hand, Markevitch et al. (1998)
found from ASCA observations that cluster temperatures decrease
significantly with radius.  Irwin \& Bregman (2000) analysed BeppoSAX
data and claimed a slight rise in cluster temperature with radius, a
result which is at odds with De Grandi \& Molendi (2002), who found
that for a set of 21 clusters the temperature profile has a clear
isothermal core (excluding the cooling flow region) followed by a
rapid radial decline. Note that such a core is consistent with Chandra
observations of cooling flow clusters, where the temperature profile
rises rapidly with radius, then remains approximately constant out to
$\sim$ 0.8~Mpc (Allen, Schmidt \& Fabian 2001). Finally, XMM-Newton
observations e.g. of Coma (Arnaud et al. 2001) and Abell 1795 (Tamura
et al. 2001) show small but significant radial variations of the
temperature.  Numerical simulations seem to confirm a decline of
temperature profiles with radius, but are not able to reproduce the
flatness of these profiles in the innermost regions (Frenk et
al. 1999; Loken et al. 2002). This disagreement may be due to
additional physical processes that must be taken into account in
future numerical simulations.  Another point is that a single $\beta$
value cannot always fit the X-ray surface brightness profile of
clusters (Allen, Ettori \& Fabian 2001; Allen, Schmidt \& Fabian 2001;
Hicks et al.  2002).

The ICM density and temperature distributions are of fundamental
importance because they can be used to determine the specific
entropy distribution of the ICM, thus providing important information
to understand nongravitational internal and external processes that
may contribute to the ICM thermal history, such as external preheating
and energy injection supernova-driven galaxy winds (Brighenti \&
Mathews 2001; Dos Santos \& Dor\'e 2002). Non-gravitational processes
may be responsible for the observed breaking of the self-similar
relation between X-ray luminosity and temperature predicted by theory
(Arnaud \& Evrard, 1999; Rosati, Borgani \& Norman 2002 and references
therein) and also the so called Entropy Floor (Ponman, Cannon \&
Navarro, 1999; Helsdon \& Ponman 2000; Lloyd-Davies et al. 2000).

With the assumption that the X-ray plasma in clusters of galaxies is
weakly dissipative, clusters considered as self-gravitating systems
are likely to verify properties similar to those recently found in
elliptical galaxies, considered as self-gravitating systems.  Namely,
the optical surface brightness profiles of elliptical galaxies can be
fit by a S\'ersic law (S\'ersic 1968; Caon et al. 1993; Ciotti \&
Bertin 1999):

\begin{equation}
\label{sersic_profile}
\Sigma(s)=\Sigma_0\:exp\left[- \left ( \frac{s}{a} \right )^{\nu} \right]
\end{equation}

\noindent
characterized by three parameters: $\Sigma_0$ (intensity), $a$
(scaling) and $\nu$ (shape). For a sample of 132 ellipticals belonging
to three galaxy clusters, the S\'ersic parameters were found to be
correlated two by two, and in the three-dimensional space defined by
these three parameters they are located on a thin line.  These
properties have been interpreted as due to the fact that, to a first
approximation, all these elliptical galaxies have the same specific
entropy (entropy per unit mass) (Gerbal et al. 1997, Lima Neto et
al. 1999, M\'arquez et al. 2000), and that a scaling law exists
between the potential energy $U$ and the mass $M$ for these galaxies:
$U\ \propto \ M^{1.72\pm0.03}$ (M\'arquez et al. 2001). Each of these
relations defines a two-manifold in the [$log \Sigma_0 , loga, \nu$]
space.  The thin line on which the galaxies are distributed in this
space is the intersection of these two two-manifolds.  Such relations
are most probably a consequence of the formation and evolution
processes undergone by these objects, since theory predicts $U\
\propto \ M^{5/3}$ under the hypothesis that energy and mass are
conserved (M\'arquez et al. 2001).

Interestingly, numerical simulations of cold dark matter haloes in two
different mass ranges lead to a similar scaling law between the
potential energy and mass of the haloes. In the mass range $4 \times 10^5\
\leq\ M\ \leq\ 4 \times 10^8\ M_\odot$ (unvirialized clusters), Jang-Condell
\& Hernquist (2001) find a relation consistent with $U\ \propto \
M^{5/3}$, while in the mass range $10^{12}\ \leq\ M\ \leq\ 10^{15}\
M_\odot$ (virialized clusters) Lanzoni (2000) finds $U\ \propto \
M^{1.69\pm0.02}$.

In this work, we present a study aimed at testing whether
results similar to those found in elliptical galaxies can also
be obtained for galaxy clusters, based on an accurate modeling of the
cluster X-ray surface brightness. We use a de-projection of the
S\'ersic profile (Eq.~\ref{sersic_profile}) to obtain the gas and
DM density distributions, temperature profiles, dynamical mass
distributions and estimations of the integrated specific
entropies of the gas and DM components for a set of 24 nearby
galaxy clusters and a group. Interesting correlations between
physical quantities are found, comparable to those observed in
elliptical galaxies, which can give important clues to understand
better the formation and evolution of galaxy clusters. This paper is
structured as follows: our sample is described in Sect.~\ref{sample};
the calculations of the physical quantities used in this paper are
presented in Sect.~\ref{theoretical_background}; the method used to
determine the gas density profile from the X-ray surface brightness is
described in Sect.~\ref{profile}; the methods used to derive the
temperature profile and the dark matter distribution are explained in
Sect.~\ref{kappa}; results are presented in Sect.~\ref{results} and
conclusions in Sect.~\ref{conclusions}.

\section{The sample}\label{sample}

We have retrieved data taken with the PSPC-B camera of ROSAT from the
ROSAT archive at MPE. The energy range considered is 0.44$-$2 keV,
corresponding to the four energy bands R4 to R7 (Snowden et al.
1994). These bands were chosen in order to avoid the low signal to
noise ratio in the lower bands due to the high absorption by the
hydrogen column. The spatial resolution of the PSPC is 25'' and its
energy resolution corresponds to 0.43\% at 0.93 keV.  We selected
observations with the longest exposure times and where the cluster
showed a regular shape, with no obvious mergers and a smooth light
curve (no strong scattered solar X-ray contamination). We thus built a
sample of 24 clusters (see Table~\ref{tab_data}) with redshifts
ranging between $0.01$ and $0.3$. Redshifts were taken from the SIMBAD
data base (except for A2199 for which the redshift was obtained from
Wu, Xue \& Fang 1999), and gas temperatures and luminosities from Wu,
Xue \& Fang (1999), except for A2034 and A2382 for which temperatures
were taken from Ebeling et al. (1996).  We assume $H_0 = 50\;\rm
km\;s^{-1}\;Mpc^{-1}$, $\Omega_0 = 1$, and $\Lambda = 0$ throughout
this analysis. In order to increase the range in $T_X$, in particular
to include cooler systems when drawing the $L_X-T_X$ relation, we
intended to add several groups to our sample of clusters. However, we
only included in our sample the group HCG~62, the one with
the longest exposure time and best signal to noise ratio (we also
tried to include HCG~94, but discarded it because of its signal to
noise ratio). More groups will be considered in forecoming works.

The data reduction was done using the software developed by Snowden et
al. (1994). The routines in the software provide the best available
modeling and subtraction of various non-cosmic background components
and corrections for exposure, satellite wobbling, vignetting and
variations of detector quantum efficiency. A flat-field correction of
the images was applied and the non-extended sources were masked,
except the cluster centers. The whole procedure was carried out only
in clusters without strong scattered solar X-ray contamination; for
each cluster, we checked the light curves in the 4 energy bands
considered, and all those with count rate peaks larger than 3 counts
sec$^{-1}$ in their light curves were excluded.

\begin{center}
\begin{table*}[h]
\centering
\caption{Cluster sample and observational data from the literature}
\begin{tabular}{llrlrr}
\hline
\rule[-2mm]{0mm}{7mm} $ Cluster$ & z & Exp. time (sec) & $T_0$ (keV) & $L_X (10^{44}$ erg s$^{-1})$ \\
\hline
A85   & 0.0518 & 10240 & $6.20^{+0.40}_{-0.15}$ & $19.52^{+1.35}_{-1.35}$ \\
A478  & 0.0881 & 21969 & $6.90^{+0.35}_{-0.35}$ & $32.00^{+4.08}_{-4.08}$ \\
A644  & 0.0704 & 10246 & $6.59^{+0.17}_{-0.17}$ & $18.92^{+2.17}_{-2.17}$ \\
A1651 & 0.0860 & 7429  & $6.10^{+0.20}_{-0.20}$ & $18.78^{+2.21}_{-2.21}$ \\
A1689 & 0.1810 & 13957 & $9.02^{+0.40}_{-0.30}$ & $55.73^{+8.92}_{-8.92}$ \\
A1795 & 0.0631 & 26172 & $5.88^{+0.14}_{-0.14}$ & $25.42^{+1.47}_{-1.47}$ \\
A2029 & 0.0765 & 12550 & $8.47^{+0.41}_{-0.36}$ & $41.93^{+2.96}_{-2.96}$ \\
A2034 & 0.1510 & 8952  & $7.00$ & $6.86$ \\
A2052 & 0.0348 & 6211  & $3.10^{+0.20}_{-0.20}$ & $4.27^{+0.34}_{-0.34}$ \\
A2142 & 0.0899 & 6186  & $9.70^{+1.30}_{-1.30}$ & $61.12^{+3.95}_{-3.95}$ \\
A2199 & 0.0299 & 40999 & $4.10^{+0.08}_{-0.08}$ & $7.09^{+0.25}_{-0.25}$ \\
A2219 & 0.2250 & 11200 & $12.40^{+0.50}_{-0.50}$ & $64.56^{+6.96}_{-6.96}$ \\
A2244 & 0.0970 & 2963  & $8.47^{+0.43}_{-0.42}$ & $25.32^{+2.14}_{-2.14}$ \\
A2319 & 0.0559 & 3169  & $9.12^{+0.15}_{-0.15}$ & $39.74^{+2.17}_{-2.17}$ \\
A2382 & 0.0648 & 17444 & $2.90$ & $0.91$ \\
A2390 & 0.2310 & 10335 & $11.10^{+1.00}_{-1.00}$ & $63.49^{+14.87}_{-14.87}$ \\
A2589 & 0.0416 & 7289  & $3.70^{+1.30}_{-0.70}$ & $3.42^{+0.38}_{-0.38}$ \\
A2597 & 0.0852 & 7163  & $4.40^{+0.40}_{-0.70}$ & $15.37^{+1.79}_{-1.79}$ \\
A2670 & 0.0761 & 17679 & $4.45^{+0.20}_{-0.20}$ & $4.97^{+0.92}_{-0.92}$ \\
A2744 & 0.3080 & 13648 & $11.00^{+0.50}_{-0.50}$ & $62.44^{+14.41}_{-14.41}$ \\
A3266 & 0.0594 & 13547 & $8.00^{+0.30}_{-0.30}$ & $16.48^{+0.64}_{-0.64}$ \\
A3667 & 0.0552 & 12560 & $7.00^{+0.60}_{-0.60}$ & $22.70^{+4.20}_{-4.20}$ \\
A3921 & 0.0960 & 11997 & $4.90^{+0.55}_{-0.55}$ & $10.92^{+1.52}_{-1.52}$ \\
A4059 & 0.0460 & 5439  & $3.97^{+0.12}_{-0.12}$ & $5.78^{+0.54}_{-0.54}$ \\
HCG62 & 0.0137 & 19639 & $1.1 ^{+0.05}_{-0.05} $  & $0.12$ \\
\hline
\end{tabular}
\label{tab_data}
\\ For A2034 and A2382, the temperatures and X-ray luminosities were taken 
from Ebeling et al. (1996) \\ who do not provide error bars.
\end{table*}
\end{center}

\section{Estimating physical quantities}\label{theoretical_background}

\subsection{Gas density profile}\label{sersic}

The observed X-ray emission of the ICM is directly related to the gas
distribution in the dark matter halo gravitational potential. Thus, in
order to compare theory with observations, a description of the gas
distribution is needed.  Using a parameter-dependent model for the gas
density profile, it is possible to re-construct the 3D X-ray emission
of the cluster which, once projected and compared to the observations
(Sect.~\ref{profile}), will allow us to derive the best set of values
for the model parameters. We have chosen a 3D deprojection of a
S\'ersic profile (S\'ersic 1968) to describe the gas distribution in
clusters. This choice was motivated by the fact that we already used
this profile to fit the optical surface brightness of elliptical
galaxies, and computed all the physical quantities needed here, such
as the entropy, potential energy, etc., as a function of the three
S\'ersic parameters (see M\'arquez et al. 2001 and references
therein). Note that from a mathematical point of view, since the
S\'ersic profile has three parameters instead of two (compared to
other models as for instance the $\beta$-model), the fitting process
is more flexible. Besides, the fact that the volume integral of this
profile does not diverge at large radii allows us to compute important
quantities such as the total mass, potential energy and entropy of the
system without any extra mathematical requirement such as a cutoff
radius, for instance. Note also that the S\'ersic law
(Eq.~(\ref{sersic_profile})) is a non-homologous generalization of the
de Vaucouleurs $R^{1/4}$ profile (de Vaucouleurs 1948). The 3D
deprojection of such a profile corresponds to a generalized form of
the Mellier-Mathez profile (Mellier \& Mathez 1987) given by:

\begin{equation}
\label{mm_profile}
\rho_{gas}(r)=\rho_0\:{(r/a)}^{-p}\:\exp[- {(r/a)}^{\nu}]
\end{equation}

\noindent
where $\rho_0$ is the volume gas density associated to the central
column density $\Sigma_0$ and the parameters $p$ and $\nu$ are related
by the numerical approximation (M\'arquez et al. 2001):

\begin{equation}
\label{p_nu}
p \simeq 1.0 - 0.6097 \nu + 0.05563 {\nu}^2
\end{equation}

\noindent
which gives the best approximation to the S\'ersic law when
Eq.~(\ref{mm_profile}) is projected. The S\'ersic profile defined by
Eq.~(\ref{sersic_profile}) corresponds to a surface mass density while
Eq.~(\ref{mm_profile}) is the volume mass density. The condition that
the mass obtained by integrating Eq.~(\ref{sersic_profile}) must be
equal to the mass obtained by integrating Eq.~(\ref{mm_profile})
implies:

\begin{equation}
\label{rho0_sigma0}
\rho_0=\frac{1}{a}\:\Sigma_0\:\frac{\Gamma(\frac{2}{\nu})}{2\:\Gamma(\frac{3-p}{\nu})}
\end{equation}
where $\Gamma(a)$ is the complete gamma function defined by $\Gamma(a) =
\int^{\infty}_0 x^{a-1} e^{-x} d x$.

\subsection{Dark matter distribution and dynamical mass}

Once the gas distribution is known, a reasonable hypothesis can be
used to derive the dark matter distribution in the cluster. Previous
works on X-ray clusters suggest a power law relation between the
distributions of dark matter and gas (e.g. Gerbal et al. 1992; Durret
et al. 1994).  We will assume here a relative distribution of the DM
and gas of the form $\rho_{DM}/\rho_{gas} = R(r)$, where $R(r)$ is a
power law of the form:

\begin{equation}
\label{ratio}
R(r) = \kappa \left(\frac{r}{a}\right)^{-\alpha} 
\label{relation_rapport}
\end{equation}

Under this hypothesis, the dark matter also follows a S\'ersic law: it
decreases exponentially above a certain radius and its asymptotic
behavior towards the cluster centre is a power law of slope
$p'=-(p+\alpha)$. The values for $\kappa$, assuming $\alpha =0.25$ are
given in Table~\ref{tab_par} (see Sect.~\ref{profile}).

Using Eqs. (\ref{ratio}) and (\ref{mm_profile}), the total amount of
mass contained within a spherical region of radius $r$ is given by the
integral:

$$M_{Dyn} (r,\kappa,\alpha) = \int^r_0\left[\kappa \left (\frac{u}{a} \right )^{-\alpha}+1\right]\rho_{gas}(u) 4 \pi u^2 \ud u =$$

\begin{equation}
\label{mass_dyn}
\frac{4 \pi \rho_0 a^3}{\nu} \left\{\kappa \ \gamma\left[\frac{3-(p+\alpha)}{\nu},\left (\frac{r}{a} \right )^{\nu}\right] + \gamma\left[\frac{3-p}{\nu},\left (\frac{r}{a} \right)^{\nu}\right]\right\}
\end{equation}

\noindent
where $\gamma(a,z)$ is the incomplete gamma function defined by
$\gamma(a,z) = \int^{z}_{0} x^{a-1} e^{-x} d x$.

\subsection{Gas temperature profile}

An important point in our study is to compute the temperature
distribution of the ICM. This can be achieved by estimating the
gas density profile, obtained by fitting the observations, and
assuming that clusters of galaxies are systems in a nearly
hydrostatical equilibrium sate. A hypothesis on the state equation of
the ICM gas is also needed. An ideal gas state equation can be
considered as a good approximation, although its application to
self-gravitating systems has been questioned in the past (see Bonnor
1956 and references therein).

Therefore, the equation of hydrostatical equilibrium:

\begin{equation}
\label{hydro_eq}
\nabla P(r) = - G \frac{M_{Dyn}(r)}{r^2} \rho_{gas}(r) 
\end{equation}

\noindent 
is then combined with the equation of state for the hot intra-cluster
plasma:

\begin{equation}
\label{state_eq}
P(r) = \frac{\rho_{gas}(r)}{\mu m_p} k_B  T_{gas}(r) \;  
\end{equation} 

\noindent
to provide the following equation from which the ICM temperature
as a function of radius $T_{gas}(r)$ can be derived once the gas
number density as a function of radius $n_{gas}(r)$ is known:

$$M_{Dyn}(r)= -\frac{k_B}{\mu m_p G} r^2 \left\{T_{gas}(r)\frac{\ud\,
\ln [n_{gas}(r)]}{\ud r}\right.+$$
\begin{equation}
\label{hydro_eq2}
+ \left.\frac{\ud T_{gas}(r)}{\ud r}\right\} \;
\end{equation}

\noindent
where $G$ is the gravitational constant, $k_B$ is the Boltzman
constant, $\mu$ is the plasma molecular weight (we assume $\mu=0.6$
for the ICM) and $m_p$ the proton mass. The electron number
density and the gas mass density are related by
$n_{gas}(r)=\rho_{gas}(r)/(1.14\ m_p)$. The gas temperature profile
can be obtained as a function of $\kappa$ and $\alpha$ by replacing
Eqs.~(\ref{mass_dyn}) and (\ref{mm_profile}) into
Eq.~(\ref{hydro_eq2}), and performing a Gauss-Laguerre integration of
the latter. The solution is of the form:

\begin{equation}
\label{temp_pro}
T(r,\kappa,\alpha) = \left(\frac{w}{\nu a}\right) \left (\frac{r}{a} \right )^p e^{{\left (\frac{r}{a} \right )}^{\nu}} t(r,\kappa,\alpha)
\end{equation}
where $w \equiv 4 \pi \, G \,\frac{\mu m_p}{k_B} \,\frac{\rho_0 a^3}{\nu} = 1.54 \times 10^{38} \left(\frac{\rho_0 a^3}{\nu}\right)\ {\rm m\;keV}$, and

$$t (r,\kappa,\alpha) = \int^{\infty}_{\left(\frac{r}{a}\right)^{\nu}} \left\{\kappa \gamma\left[\frac{3-(p+\alpha)}{\nu},x\right]\right.+$$

\begin{equation}
\label{int_temp}
+ \left.\gamma\left[\frac{3-p}{\nu},x\right] \right\} x^{-\frac{(p + \nu + 1)}{\nu}} e^{-x} \ud x
\end{equation}

Eq.~(\ref{temp_pro}) will be used together with a model for the hot
plasma spectrum to derive the best set of values for $\kappa$ and
$\alpha$, in agreement with quantities already observed in clusters of
galaxies such as their mean X-ray temperature and luminosity and their
baryonic mass fraction (see Sect.~\ref{kappa}).

\subsection{Potential energy and specific entropy}\label{entropy}

Changing the upper limit of the integral in Eq.~(\ref{mass_dyn}) to
$\infty$ we obtain the total dynamical mass:

\begin{equation}
\label{dyn_mass}
M_{Dyn}= \frac{4 \pi \rho_0 a^3}{\nu}\left\{\kappa \, \Gamma\left[\frac{3-(p+\alpha)}{\nu}\right]+\Gamma\left(\frac{3-p}{\nu}\right)\right\}
\end{equation}
 
The total potential energy of the cluster is given by:
 
\begin{equation}
\label{u}
U_{pot} = G \frac{M^2_{Dyn}}{R_g}
\end{equation}

\noindent
where the gravitational radius $R_g$ is defined by
$R_g\:=\:a\:R^*_g$, where $a$ is the scale parameter and $R^*_g$ is a
dimensionless radius given by the numerical approximation:
$$ln(R^*_g)\:\simeq\:\frac{0.82032-0.92446\:\ln(\nu)}{\nu}\:+\:0.84543\;\;$$
\noindent 
(M\'arquez et al. 2001).

In spite of the X-ray emission, which is responsible in many cases for
cooling flow processes affecting the equilibrium state of the cluster
in the inner regions, we may consider clusters as structures where
dissipating processes are negligible compared to their gravitational
energy, thus settling into a nearly thermodynamic equilibrium at large
scale. This can be inferred by a simple order of magnitude
calculation: the potential energy of a cluster of mass $\sim 10^{15}
M_\odot$ and radius $\sim$ 1~Mpc is about $8 \times 10^{64}$ ergs. The
energy lost through X-ray emission during a Hubble time ($ \sim 2
\times 10^{17}$~s) by a cluster of X-ray luminosity $ \sim 10^{45}$
erg~s$^{-1}$ is around $3 \times 10^{62}$~ergs, a value which is
almost 300 times smaller than its potential energy.  We can therefore
estimate the gas entropy of such a configuration.

The specific entropy of the intra-cluster gas can be computed from the
distribution function in the phase space, $f(\mathbf{x},\mathbf{v})$,
of the gas particles using the microscopic Boltzmann-Gibbs definition:

\begin{equation}
\label{boltzmann-gibbs-entropy}
s = - \frac{\int f\:ln (f)\:d^3x\:d^3v}{\int f\:d^3x\:d^3v} 
\end{equation}

\noindent
where the Boltzmann constant is $k_B=1$. Note that this expresion
gives us the specific entropy of the entire system (gas or DM) because
the integration covers the total volume in phase space. This
definition then corresponds to an integrated specific entropy, and it
will be referred to as 'global' specific entropy. It is important to
say that when we use the words 'integrated' or 'global' for the
specific entropy, we are referring to this definition applied to the
gas or DM separately and not to the sum of these two components.

To find the distribution function some simplifying hypotheses are
needed.  The first one is that our system is spherically symmetric,
and the second one is that the velocity dispersion of the gas
particles is isotropic (we neglect any possible rotation of the
gas). Thus the distribution function $f$, depending explicitly
only on the total energy, can be obtained from the density
profile by an Abel inversion (Binney \& Tremaine 1987). In this way,
$f=f(\rho)$ and the integrated specific entropy can be computed
numerically as a function of the S\'ersic parameters only, providing a
way to estimate the integrated specific entropy of the ICM from its
surface brightness fit. It is important to emphasize here that
the Boltzmann-Gibbs definition of the specific entropy makes no
assumption on the equation of state, in particular the ideal gas
equation for the ICM, and no assumption either on the validity of a
hydrostatic equilibrium state of the system; this definition is
therefore a general one since it only assumes spherical symmetry and
an isotropic velocity dispersion tensor.  Since the method we present
here, based on the cluster X-ray surface brightness fitting, provides
a good constrain on the gas density profile, this density profile,
together with Eq.~(\ref{boltzmann-gibbs-entropy}) can be used to
obtain a good estimate of the global ICM specific entropy.

The specific entropy can also be obtained from the following set of
equations (Balogh, Babul \& Patton 1999):

\begin{equation}
\label{sent_prof}
s(r)= c_v \left\{ ln \left [ \frac{2 \pi (\mu m_p)^{8/3} K_0(r)}{h^2}  \right ] + \frac{5}{3} \right\}\;  
\end{equation}

\noindent
and

\begin{equation}
\label{sent_prof_k}
K_0(r) = \frac{k_B T_{gas}(r)}{\mu m_p \rho_{gas}(r)^{2/3}}\;  
\end{equation}

\noindent
where $c_v$ is the specific heat capacity at constant volume of
the plasma. These widely used equations require, however, the
following assumptions on the ICM : the gas is considered to be
monoatomic and the equation of state for an ideal gas is supposed to
hold. By knowing the ICM density and temperature distributions,
Eqs.~(\ref{sent_prof}) and (\ref{sent_prof_k}) can then be used (see
Sect.~\ref{specific_entropy}) to compute the gas specific entropy
profile for each cluster. Gas density profiles are obtained from the
X-ray surface brightness fit (see Sect.~\ref{profile}) while
temperature profiles are derived by assuming both the hydrostatic
equilibrium condition and the ideal gas state equation for the ICM, as
explained before. Since the physical quantities involved here are
intensive, these equations cannot be integrated to derive the integrated
specific entropy of the gas, and Eq.~(\ref{boltzmann-gibbs-entropy})
will have to be used for this purpose.

The entropy is of fundamental importance to understand the
effects of non-gravitational processes on the thermodynamical history
of clusters. Previous studies refer to the gas entropy at
$0.1r_{vir}$, where $r_{vir}$ is the cluster virial radius
(e.g. Ponman, Cannon \& Navarro, 1999; Lloyd-Davies, Ponman \& Cannon
2000), while in this work we estimate the gas specific entropy
for the entire cluster. Our calculations take into account all
possible sources of heating and cooling, regardless of the distance to
the centre, thus providing a good quantitative basis to which models
can be compared in order to disentangle the different processes that
affect the internal energy of the ICM during its history since the
earlier epochs of the cluster formation.

We also notice that our assumption concerning the
$\rho_{DM}/\rho_{gas}$ ratio (Eq.~(\ref{ratio})) implies that the
resulting DM density distribution is also S\'ersic-like, with a
central density distribution described by a power law varying as
$r^{-(p+\alpha)}$. Under the hypothesis already stated concerning the
distribution function of particles in phase space, we can in principle
also compute the integrated specific entropy of the DM halo by using
Eq.~(\ref{boltzmann-gibbs-entropy}) and the DM density profile.
Therefore the DM specific entropy is calculated numerically from the
equation:

\begin{equation}
\label{sent_dm}
s=-\frac{1}{M}\int^{\Psi(0)}_0 \frac{dM}{d\epsilon} ln f(\epsilon) d\epsilon\;,
\end{equation}

\noindent
which can be derived from Eq.~(\ref{boltzmann-gibbs-entropy}), and
where $\epsilon$ represents the binding energy, $M$ the total halo
mass, and $\Psi$ the relative gravitational potential (see Magnard 2002).

\section{Fitting method and density profile}\label{profile}

Finding the gas density profile amounts to deriving the best set of
values for the S\'ersic parameters. This has required to fit the ROSAT
images by a pixel-to-pixel method, which creates a three-dimensional
model of the X-ray emission which is then projected by integration
along the line of sight, taking into account the energy response and
the point spread function of the detector. The result is a synthetical
image which can be compared to the observation, and the best set of
S\'ersic parameters is obtained by successive iterations.  The gas
density profile defined by Eq.~(\ref{mm_profile}) is used to model the
bremsstrahlung emission, and the free-bound and bound-bound X-ray
emissions are taken into account. The code computes the X-ray
emissivity $\epsilon_{\nu}$ in every point; it is then projected to
obtain the surface brightness.  This projected image is then convolved
with the ROSAT point spread function (PSF), which varies as a function
of position (and energy) on the detector. We have used in our fits a
FWHM of 2 pixels which corresponds to the central PSPC PSF.

The cluster redshift and the mean gas temperature are required as
input parameters. To obtain an initial guess for the free parameters
in Eq.  (\ref{mm_profile}) we fit a S\'ersic profile (Eq.
(\ref{sersic_profile})) to the X-ray surface brightness of each
cluster. In order to stay as close as possible to our hypothesis of
spherical symmetry, we selected clusters presenting the most round and
uniform projected X-ray emission. However, this emission is not
perfectly circular and due also to the fact that what we observe is
always a projection on the plane of the sky of a three dimensional
structure, we decided to take into account during the fitting process
the ellipticity of the projected emission supposing also that all the
clusters are oblate spheroids, i.e., that the third axis along the line of
sight is a major axis. If $a$ is the semi--major axis, which also
corresponds to the scale parameter in the S\'ersic profile, and $b$ is
the semi--minor axis, we define the ellipticity of the projected X-ray
emission by $e=\sqrt{1-(b/a)^2}$. The ellipticity and semi-major axis
position angle of the X-ray distribution are thus considered as free
parameters for the fit and given as inputs for the code. The gas was
assumed to be isothermal as a first approximation.

\vspace{0.5cm}
\begin{center}
\begin{figure}[htbp]
\centerline{\psfig{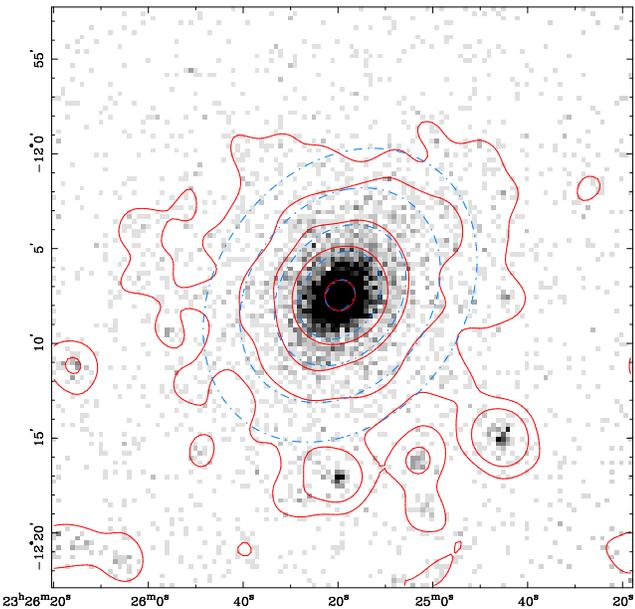}}
\caption{Comparison of our best model of the X-ray surface brightness
(dot-dashed ellipses) with the corresponding ROSAT image for the
cluster A2597. The X-ray iso-contours are 0.5, 1, 3, 10 and 100 times
the background; they were made after smoothing the original image with
a gaussian kernel of $\sigma = 3.2$ pixels.}
\label{model}
\end{figure}
\end{center}

After obtaining the initial guesses for each free parameter, we use
these values together with Eq.~(\ref{mm_profile}) to make a new
synthetic image, then compare it to the actual ROSAT image. The
parameter values are then changed and the comparison process is
repeated iteratively, until it finds the 3D X-ray emission which best
fits the surface brightness profile of the observation when
projected. The fitting process is carried out with the MIGRAD method
in the CERN MINUIT library (James 1994). In this process, the gas
temperature is kept constant, as a first approximation. The exact
shape of the temperature profile is not crucial when finding the set
of values for the parameters, because the emissivity does not have a
strong dependence on temperature as compared to its dependence on
density; indeed, we find that modifying the temperature profile
changes the values of the S\'ersic parameters by at most a few
percent, as explained in Sect.~\ref{kappa}.

The fitting process directly gives us the best estimates for the
semimajor axis $a$ and shape parameter $\nu$. However, the central
electronic number density ${n_e}_0$ given by the fit is not accurate
enough, and is estimated by another method (see Sect.~\ref{kappa}).

We evaluated errors with the MINUIT error function MINOS, which
calculates parameter errors taking into account both parameter
correlations and non-linearities.  Resulting errors correspond to a
$1\sigma$ deviation.

All the equations presented in Sect.~\ref{theoretical_background}
refer to spherical symmetry, for which we had to define an equivalent
scale parameter, $a_{eq}$ in order to go from the oblate geometry
considered in the fit to the spherical geometry of the model. This new
scale is defined as $a_{eq} = (a^2 b)^{1/3}$ and will be used instead
of $a$ to compute the specific entropy, the dynamical mass and the
potential energy of a spherically symmetric X-ray region.

We show in Fig.~\ref{model} a comparison of our best fit S\'ersic
model of A2597. The good fit of the cluster surface brightness
can clearly be seen, confirming the capability of the S\'ersic profile
to reproduce the cluster X-ray emission. 

\section{Gas temperature and dark matter to gas ratio}\label{kappa}

The fitting process described above provides the best values of
$a_{eq}$ and $\nu$ for a given gas temperature profile, assuming as a
first approximation that each cluster has an isothermal ICM, and then
obtaining, through an iterative procedure, the best compatible gas
density and non-isothermal temperature profiles, assuming hydrostatic
equilibrium and spherical symmetry.

In order for the ${n_e}_0$ normalization to give the observed number
of counts, we first computed the cluster flux $F$, after substraction
of point sources and background (obtained by the 2D fit). The masked
source pixel counts were set to the mean value taken within ellipses.
Then ${n_e}_0$ was chosen so that a bremsstrahlung XSPEC model
convolved by a hydrogen column absorption would give the same flux as
seen by ROSAT:

$$F=\frac{3.02\times10^{-15}}{4 \pi d^2_L}\int n_e n_I dV = $$

$$\frac{3.02\times10^{-15}}{4 \pi d^2_L}\ \frac{2^{2+\frac{2p-3}{\nu}}}{\nu} a^3 {{n_e}_0}^2 \pi \Gamma\left[{\frac{3-2p}{\nu}}\right]$$

\noindent
where $d_L$ is the luminosity distance to the cluster, and $n_I$
is the number density of plasma ions.

Eqs. (\ref{temp_pro}) and (\ref{int_temp}) provide the ICM temperature
profiles as families of solutions depending on the parameters $\kappa$
and $\alpha$. Since these are found to be different from those assumed
to estimate the S\'ersic parameters, the fitting process must be
repeated, using the new temperature profile, in order to obtain a new
set of parameters for each cluster, until convergence.

We tested on one cluster (A2029) the effect of a temperature gradient
on the gas density profile from distances close to $r_{eff}/4$
outwards, where $r_{eff}$ corresponds to an effective radius which
contains half of the cluster gas mass (see Table~\ref{tab_par}). For
this, we did a second fit of the surface brightness of A2029 using a
new temperature profile given by a power law of the form $T(r)=T_0
(r/r_{eff})^\beta$ where $r_{eff} \simeq 2800$~kpc. $T_0$ was set in
order to have a non-weighted mean temperature equal to the mean
cluster temperature (Table~\ref{tab_data}) and we considered the cases
$\beta = -0.5$ and $-1$. After the first iteration, the values found
for ${n_e}_0$, $a_{eq}$, and $\nu$ remained almost unchanged for both
values of $\beta$: the scale and shape parameters changed by about
1\%, and the central electronic density by 4\% with respect to the
isothermal fit.

We therefore decided to keep the S\'ersic values given by the original
fit as the definitive ones. Eq.~(\ref{temp_pro}) can be re-written in
the form $T(r,\kappa,\alpha) = \kappa T_1(r,\alpha)+ T_2(r)$. We
produced a set of $T_1(r,\alpha)$ profiles corresponding to values of
$\alpha$ between 0 and 2 with steps of $\Delta \alpha=0.1$. These
curves together with their corresponding $T_2(r)$ profiles and an
adequate set of values for $\kappa$ were used to find which
combination of $\kappa$ and $\alpha$ gives the closest values to the
observed global temperatures and luminosities (Ebeling et al. 1996;
Wu, Xue \& Fang 1999), and baryon mass fractions (Arnaud \& Evrard
1999; Mohr, Mathiesen \& Evrard 1999; Schindler 1999). These
quantities are evaluated as follows. The two profiles $n_e(r)$ and
$T(r,\kappa,\alpha)$ plus the hypothesis of a gas metallicity equal to
$0.3\, Z_\odot$ (see Renzini 1997) and the emission process determine
the cluster X-ray emission.  We used the XSPEC software to simulate
the corresponding spectra with a bremsstrahlung emission model (to be
coherent with the emission model used in the fitting program).  We
summed up the spectra from shells of constant electronic density and
temperature over radius, considering only the volume intersected by
the ROSAT observation cylinder. The spectrum obtained is then
convolved with a photoelectric absorption model and with the ROSAT
response function to produce the simulated spectrum.  A fit is then
performed on this spectrum to derive the X-ray temperature and
luminosity (the metallicity and hydrogen column density are fixed).

The set of $\kappa$ and $\alpha$ has to produce $L_X$ and $T$ in
agreement with the observed data within errors. Moreover, the gas mass
fractions should stay inside the observed limits.  So we performed a
minimization of the distance between the observed data and the
predictions from our model. We found that the best $\alpha$ is often
very close to zero, which is not in good agreement with results
obtained from numerical simulations by Teyssier (2002) which show a
$\rho_{DM}/\rho_g$ ratio varying globally as $r^{-0.25}$; besides our
values of $\alpha$ are not well constrained. We therefore chose to
impose $\alpha =-0.25$; $\kappa$ was then recomputed to give
luminosities and temperatures as close as possible to the
observations.

To each temperature profile $T(r,\kappa,\alpha)$ corresponds a couple
of simulated observational parameters $(T_{\rm sim},L_{\rm sim})$.
The value of $\kappa$ is constrained by imposing these parameters to
be close (within error bars) to the real observational values.  The
gas mass fraction is checked to be compatible with the isothermal
hypothesis.

\section{Results}\label{results}

\subsection{Mass distributions and parameter correlations}\label{correlations}

\begin{center}
\begin{figure}[htbp]
\centerline{\psfig{file=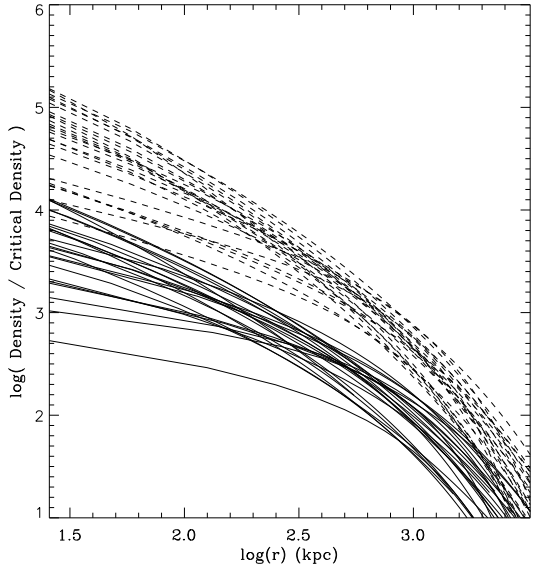,width=8cm}}
\caption{3D deprojected gas (full lines) and dark matter (dashed lines)
density profiles divided by the critical density of the Universe at the cluster redshift.}
\label{dens_profs}
\end{figure}
\end{center}

The 3D gas density profiles were computed with Eq.~(\ref{mm_profile})
for all the clusters in our sample, using the sets of parameters
obtained from the surface brightness fitting process (described in
Sect.~\ref{profile}) of the PSPC images and listed in
Table~\ref{tab_par}. By means of Eq.~(\ref{ratio}), the corresponding
DM distribution can be recovered. In Fig.~\ref{dens_profs} we show the
3D gas and DM density profiles for every cluster divided by the
critical density of the Universe at the cluster redshift (e.g.,
Ettori, De Grandi \& Molendi 2002): $\rho_c(z)=3 H^2(z)/(8 \pi G)$,
where the Hubble constant at redshift $z$ is defined by $H(z)=H_0
\sqrt{\Omega_m (1+z)^3+1-\Omega_m}$.

\vspace{0.5cm}

Both DM and gas distributions are S\'ersic-like, implying that the
corresponding profiles decrease exponentially outwards above a certain
distance from the cluster centre; on the other hand, when $r$ goes to
zero they follow a power law in radius with a logarithmic slope
tending asymptotically to $-p(\nu)$ for the gas and to
$-(p(\nu)+\alpha)$ for the DM.

The values for $\kappa$, obtained as described in Sect.~\ref{kappa},
are in most cases smaller than 10, while the value for $\alpha$ was
fixed to 0.25 (see above). Although DM halos are denser than the ICM
gas, as implied by the $\kappa$ factor and the asymptotic slope
difference $\alpha$, the general shapes of both profiles look quite
similar, implying that dark matter and gas are distributed in a
comparable way. This effect is a natural consequence of the hypotheses
and conditions imposed to our model. However such a behaviour seems to
be the case for massive systems (like those in our sample~- see total
masses in Table~\ref{tab_phy_par}), as shown by observations of galaxy
clusters at moderate and high redshifts (Schindler 1999). The gas
would be accreted into the forming structure and once the system
reaches a relaxed state, the gas just accommodates into the halo
potential. This is true at the scale of massive galaxy clusters where
the potential well is deep enough to prevent the gas from expansion
due to non-gravothermal processes. In this case, both dark matter and
gas present similar distributions in contrast with what is observed in
smaller systems such as groups of galaxies and even galaxies, where
the gas can produce extended cores as result of energy injection due
to supernova explosions or shock winds (Bryan 2000, Bower et al. 2001,
Brighenti \& Mathews 2001). It is also important to mention that
Eq.~(\ref{ratio}), based on galaxy cluster observations, may no longer
be valid for low mass systems such as groups of galaxies, in which
case our DM model would be inappropriate to describe groups.

The averaged gas density profile for our set of 24 galaxy clusters is
well fit by a S\'ersic profile with parameters: $\rho_0 = 7.4 \times
10^{-24}$~kg~m$^{-3}$, $a_{eq}= 367$~kpc and $\nu= 0.56$. The latter
gives $p(\nu)=0.67$ which makes the corresponding DM density profile
vary as $r^{-0.92}$ when $r$ goes to zero. This central slope for the
DM halo is shallower than the self-similar solution for spherical
collapse in an expanding universe found by Bertschinger (1985) ($\rho
\propto r^{-2.25}$), than the asymptotic behaviour found by Moore et
al. (1999) in their numerical simulations ($\rho \propto r^{-1.5}$)
and than the NFW (Navarro, Frenk \& White, 1996, 1997) universal
density profile.  However it is steeper than the $\rho \propto
r^{-0.75}$ critical solution found by Taylor \& Navarro (2001) for
galaxy-sized CDM halos based on the study of their phase-space density
distribution. This critical solution can be interpreted as a maximally
``mixed'' configuration, where the phase-space distribution across the
system is the most uniform one compatible with a monotonically
decreasing density profile and with the power-law entropy
distribution. This configuration would be the result of
non-spherically symmetric formation processes through hierarchical
merging. Mass shells are continously mixed and the density profiles
tend to be shallower than the NFW profile at the center, converging to
the $\rho \propto r^{-0.75}$ distribution for maximal mixing (Taylor
\& Navarro 2001).

One important advantage of the S\'ersic density profile is that its
volume integral converges when integrated up to infinity, making it
thus possible to determine the total cluster mass, and the potential
energy and entropy of the system without introducing a cutoff radius,
in contrast to the popular $\beta$-model, for instance. Total
dynamical masses can thus be computed (see Eq.~(\ref{dyn_mass})) and
the resulting values for all clusters are given in
Table~\ref{tab_phy_par}.

\vspace{0.5cm}
\begin{center}
\begin{figure}[htbp]
\centerline{\psfig{file=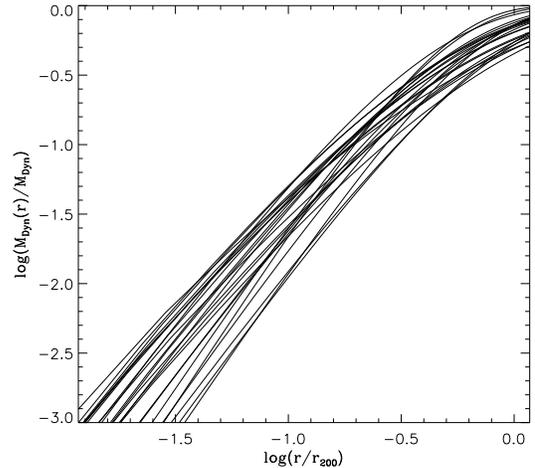,width=8cm}}
\caption{Synthetic cumulative dynamical mass profiles normalized to
the corresponding total integrated cluster mass $M_{Dyn}$. The radial
coordinate is normalized to the cluster $r_{200}$ radius.}
\label{mass_profs}
\end{figure}
\end{center}
 
Fig.~\ref{mass_profs} shows the resulting cumulative dynamical mass
profiles for every cluster as obtained by means of Eq.
(\ref{mass_dyn}) with the corresponding S\'ersic parameters.  Every
profile has been normalized to the corresponding cluster total mass,
$M_{dyn}$ (see Eq.~(\ref{dyn_mass})), and the radial coordinate has
been normalized to the corresponding cluster $r_{200}$ radius, which
is defined as the radius within which the mean density is $200$ times
the critical density of the universe, $\rho_c$, as defined above. In
general, we can define a radius $r_{\Delta}$ within which the mean
density is $\Delta$ times $\rho_c$ and $r_\Delta$ can be obtained from
the relation $M(r < r_{\Delta})/(4 \pi r^3_{\Delta}/3)=\Delta
\rho_c$). The averaged cumulative dynamical mass distributions show a
logarithmic slope $d log M / d log r \sim 1$ at $r \sim 0.7
r_{200}$. Further out this slope decreases rather fast.  According to
our model, the cumulative mass profiles converge only around $r \sim
10 r_{200}$, but the mass variation is only of a factor of two, going
from $M \sim 10^{15}\ M_{\odot}$ at $\sim r_{200}$ to $M \sim 2 \times
10^{15}\ M_{\odot}$ at $\sim 10 r_{200}$.

The analysis by Ettori, De Grandi \& Molendi (2002) of BeppoSAX data
includes some clusters of our sample. They estimate masses for $\Delta
=$ 1000 and 2500, by using either a NFW or a King profile for the
total mass distribution. Comparing our results with theirs for same
clusters, we see that our mass estimates, derived from the S\'ersic
profile for these two values of $\Delta$ within same $r_{\Delta}$
radii as indicated in their Table~2 are in quite good agreement. In
most cases, our masses differ by about 10\% or less, with differences
reaching about 20\% for a few cases. These differences are likely to
be due to the use of different functional forms (the S\'ersic, NFW and
King models) to describe the mass profile. Moreover, from our sample
we find in average $r_{200} \sim 2.5 \pm 0.4$ Mpc (see
Table~\ref{tab_par}). These values are about 78\% of those found in
numerical simulations by Navarro, Frenk \& White 1996 for cluster
sized systems of comparable masses. Moreover, we note that the values
of our scale parameter $a_{eq}$ are comparable to those of the scale
radius $r_s$ of the NFW profile for similar mass ranges. In this way,
we obtain in average $a_{eq}/r_{200} \sim 0.12$ in good agreement with
the value of $r_s/r_{200} \sim 0.14$ (Navarro, Frenk \& White 1996),
the difference being of only 14\%. We can therefore say that the
S\'ersic profile is able to describe the mass distribution in clusters
in as much the same way as other models as the NFW and King models,
the differences being due to the mathematical natures of the models.

\begin{center}
\begin{figure}[h]
\centerline{\psfig{file=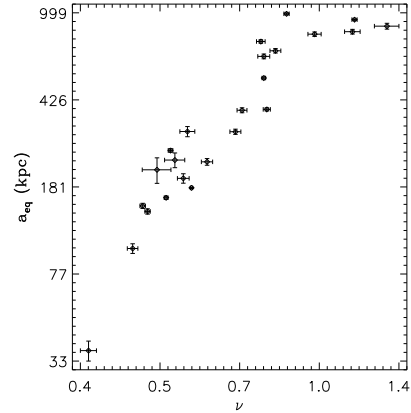,width=6cm}}
\caption{Correlation between the density profile parameters $a_{eq}$ and $\nu$.}
\label{corr_a_nu}
\end{figure}
\end{center}

\begin{center}
\begin{figure}[h]
\centerline{\psfig{file=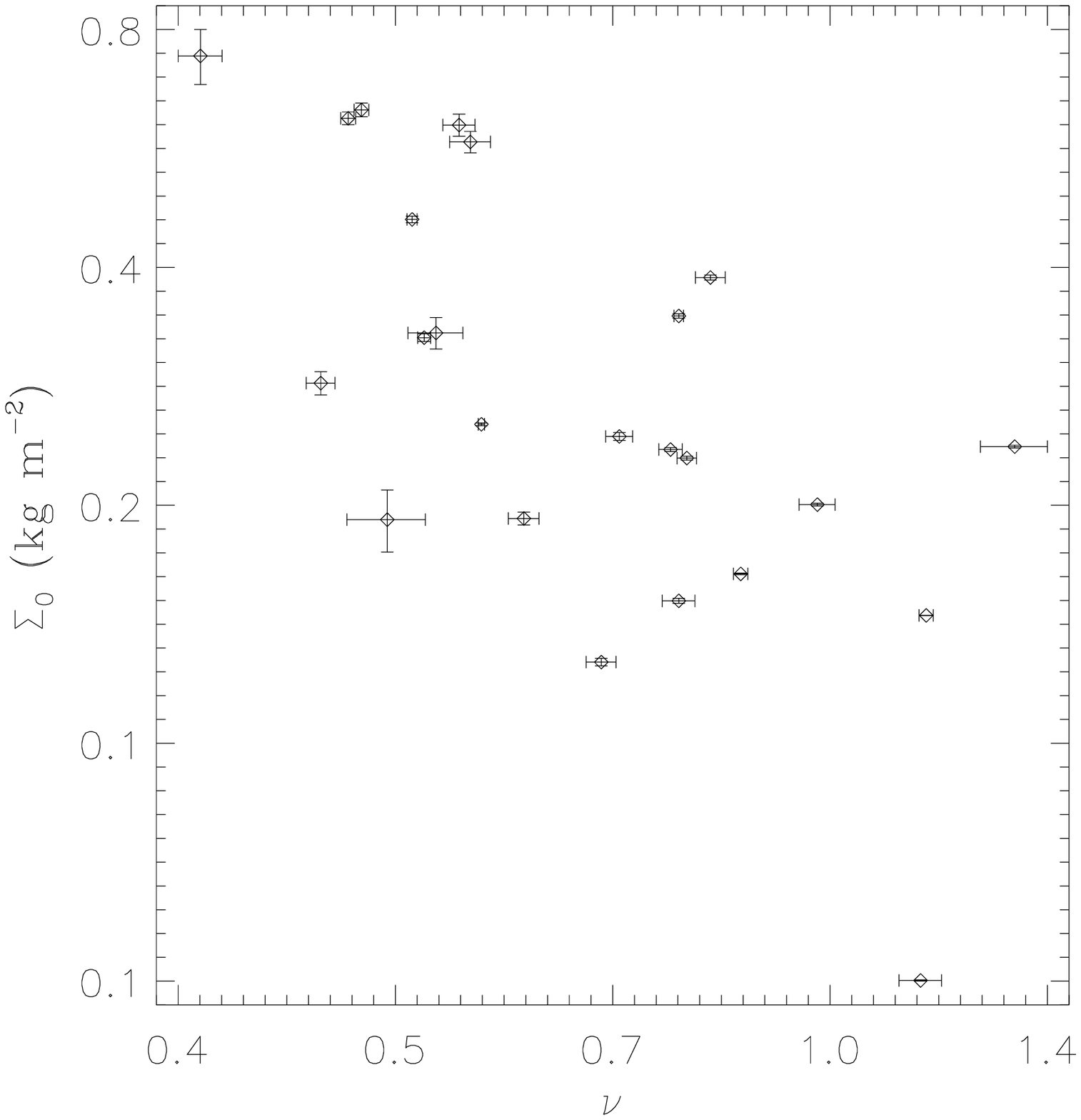,width=6cm}}
\caption{Correlation between the density profile parameters $\Sigma_0$ and $\nu$.}
\label{corr_s0_nu}
\end{figure}
\end{center}

\begin{center}
\begin{figure}[h]
\centerline{\psfig{file=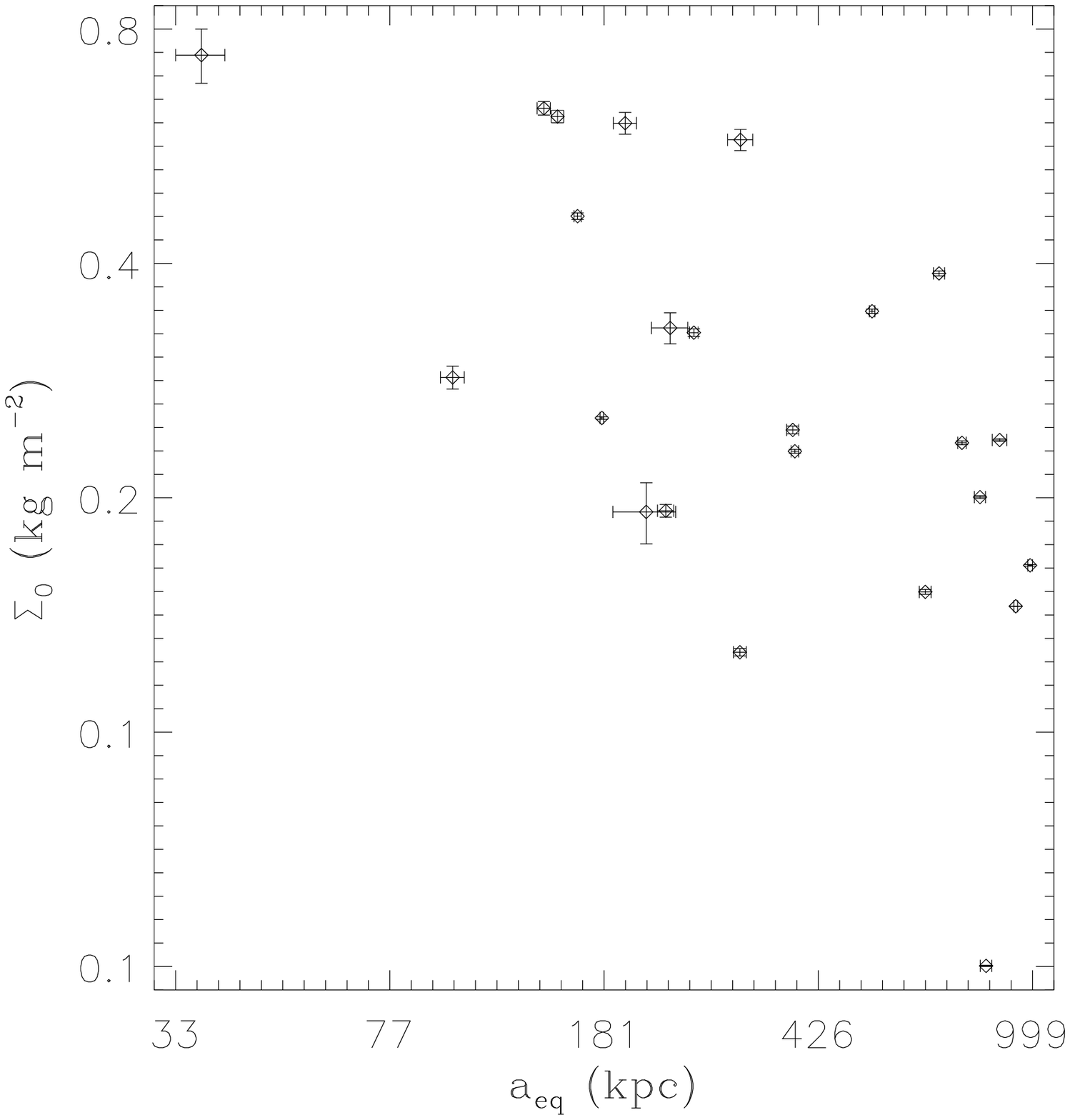,width=6cm}}
\caption{Correlation between the density profile parameters $\Sigma_0$ and $a_{eq}$.}
\label{corr_s0_a}
\end{figure}
\end{center}

Based on a de-projected S\'ersic model for the gas density profile,
the best set of values of ${n_e}_0$, $a_{eq}$ and $\nu$ for each
cluster is given in Table~\ref{tab_par}.  These three parameters are
displayed two by two in Figs. \ref{corr_a_nu}, \ref{corr_s0_nu} and
\ref{corr_s0_a}. A clear correlation between the shape ($\nu$) and
scale ($a_{eq}$) parameters is seen, while the correlations we find
for $\Sigma_0-\nu$ and $\Sigma_0-a_{eq}$ ($\Sigma_0$ being obtained
from $\rho_0$ by using Eq.~(\ref{rho0_sigma0})), are clear but show a
somewhat larger scatter. Note that these three correlations have
shapes similar to those found in elliptical galaxies. This may
indicate, as for ellipticals, the existence of an entropic line on
which galaxy clusters lie, in which case the correlation shown in
Fig.~\ref{corr_a_nu} may just be the projection of this isentropic
relation on the $log (\nu) - log (a_{eq})$ plane. In the case of
elliptical galaxies, this entropic line was interpreted as the
intersection of two surfaces in the [log $\Sigma_0$, log a, log $\nu$]
space : the Entropic surface and the Energy-Mass surface (M\'arquez et
al. 2001). In this work, we find that these two surfaces exist for our
galaxy clusters as well, and these clusters are also located on a line
corresponding to the intersections of the two surfaces (see
Sect. \ref{discussion}). A complete discussion on this point will be
presented in a forecoming paper (Magnard, in preparation). It is worth
noting that, although from the mathematical point of view no
correlation between the model parameters is expected, the fact that we
observe such correlations probably indicates that they are due to the
physics underlying the X-ray emission distribution.

\begin{center}
\begin{table*}
\caption{Best fit values for the Sersic parameters of the ICM, $\kappa$ from  
Eq.~(\ref{ratio}), and radius $r_{200}$ at which the mean density is 200 times the critical density of the Universe.}
\centering
\begin{tabular}{p{1.5cm}*{5}{c}}
\hline
\rule[-2mm]{0mm}{7mm}Cluster & $ \nu $ & $a_{eq}$ (kpc) & ${n_e}_0 \times 10^{-3}$ (cm$^{-3})$ & $ \kappa $ & $r_{200}$ (kpc) \\
\hline
\rule[-0mm]{0mm}{4mm}A85   & 0.55$\pm$0.01 & 260$\pm$5  & 5.72$\pm$0.11   & 7.17$\pm$0.32     & 2714 \\
\rule[-0mm]{0mm}{4mm}A478  & 0.50$\pm$0.01 & 143$\pm$4  & 18.51$\pm$0.52  & 7.04$\pm$0.42     & 2607 \\
\rule[-0mm]{0mm}{4mm}A644  & 0.82$\pm$0.01 & 389$\pm$6  & 3.58$\pm$0.06   & 7.19$\pm$0.21     & 2421 \\
\rule[-0mm]{0mm}{4mm}A1651 & 0.74$\pm$0.02 & 385$\pm$9  & 3.57$\pm$0.11   & 6.59$\pm$0.25     & 2453 \\
\rule[-0mm]{0mm}{4mm}A1689 & 0.58$\pm$0.01 & 198$\pm$9  & 14.26$\pm$0.82  & 7.90$\pm$0.35     & 2594 \\
\rule[-0mm]{0mm}{4mm}A1795 & 0.54$\pm$0.00 & 164$\pm$2  & 12.30$\pm$0.23  & 7.64$\pm$0.21     & 2462 \\
\rule[-0mm]{0mm}{4mm}A2029 & 0.49$\pm$0.01 & 151$\pm$4  & 16.63$\pm$0.52  & 8.65$\pm$0.43     & 2946 \\
\rule[-0mm]{0mm}{4mm}A2034 & 1.00$\pm$0.03 & 811$\pm$18 & 1.72$\pm$0.04   & 3.58$\pm$0.43     & 2450 \\
\rule[-0mm]{0mm}{4mm}A2052 & 0.47$\pm$0.01 & 100$\pm$5  & 11.70$\pm$0.72  & 9.90$\pm$0.67     & 1930 \\
\rule[-0mm]{0mm}{4mm}A2142 & 0.81$\pm$0.01 & 528$\pm$6  & 3.86$\pm$0.05   & 4.26$\pm$0.50     & 2824 \\
\rule[-0mm]{0mm}{4mm}A2199 & 0.60$\pm$0.00 & 180$\pm$1  & 6.82$\pm$0.06   & 9.42$\pm$0.20     & 2142 \\
\rule[-0mm]{0mm}{4mm}A2219 & 0.85$\pm$0.02 & 689$\pm$15 & 3.40$\pm$0.08   & 4.04$\pm$0.24     & 2887 \\
\rule[-0mm]{0mm}{4mm}A2244 & 0.56$\pm$0.02 & 237$\pm$17 & 6.38$\pm$0.68   & 13.05$\pm$0.70    & 3039 \\
\rule[-0mm]{0mm}{4mm}A2319 & 0.80$\pm$0.01 & 755$\pm$13 & 1.86$\pm$0.04   & 5.13$\pm$0.10     & 3294 \\
\rule[-0mm]{0mm}{4mm}A2382 & 1.17$\pm$0.04 & 831$\pm$19 & 0.49$\pm$0.01   & 6.61$\pm$0.66     & 1900 \\
\rule[-0mm]{0mm}{4mm}A2390 & 0.59$\pm$0.02 & 313$\pm$16 & 8.61$\pm$0.48   & 5.13$\pm$0.62     & 2658 \\
\rule[-0mm]{0mm}{4mm}A2589 & 0.72$\pm$0.02 & 312$\pm$8  & 2.31$\pm$0.07   & 9.25$\pm$1.73     & 2008 \\
\rule[-0mm]{0mm}{4mm}A2597 & 0.39$\pm$0.01 & 37$\pm$4   & 71.44$\pm$46.76 & 12.42$\pm$1.60    & 2035 \\
\rule[-0mm]{0mm}{4mm}A2670 & 0.52$\pm$0.03 & 215$\pm$27 & 4.05$\pm$0.74   & 11.44$\pm$0.58    & 2331 \\
\rule[-0mm]{0mm}{4mm}A2744 & 1.35$\pm$0.07 & 877$\pm$25 & 2.26$\pm$0.07   & 4.82$\pm$0.29     & 2506 \\
\rule[-0mm]{0mm}{4mm}A3266 & 1.18$\pm$0.01 & 935$\pm$7  & 1.22$\pm$0.01   & 5.66$\pm$0.20     & 2973 \\
\rule[-0mm]{0mm}{4mm}A3667 & 0.89$\pm$0.01 & 990$\pm$9  & 1.07$\pm$0.01   & 3.29$\pm$0.36     & 2723 \\
\rule[-0mm]{0mm}{4mm}A3921 & 0.81$\pm$0.02 & 653$\pm$15 & 1.41$\pm$0.04   & 3.66$\pm$0.50     & 2090 \\
\rule[-0mm]{0mm}{4mm}A4059 & 0.64$\pm$0.01 & 233$\pm$8  & 4.34$\pm$0.18   & 8.46$\pm$1.02     & 2033 \\
\rule[-0mm]{0mm}{4mm}HCG62 & 0.36    & 22   & 17.4      & 36.9$\pm$1.63     & $--$ \\
\hline
\end{tabular}
\label{tab_par}
\end{table*}
\end{center}

\subsection{Temperature profiles} \label{temp_prof}

\begin{center}
\begin{figure}[h]
\centerline{\psfig{file=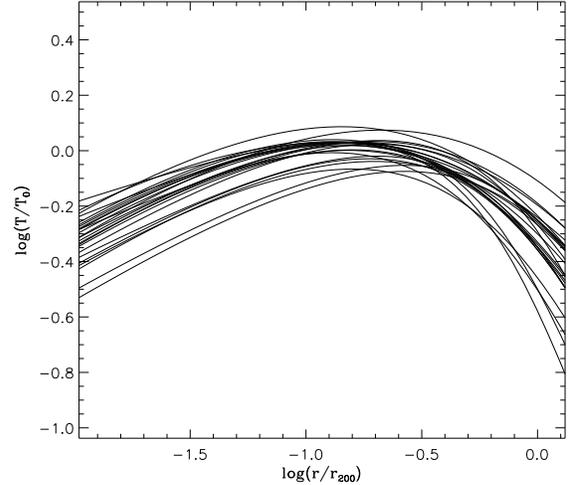,width=8cm}}
\caption{Synthetic 3 D temperature profiles calculated from
Eqs. (\ref{temp_pro}) and (\ref{int_temp}) with the values of $\alpha$
and $\kappa$ given in Table~\ref{tab_par}. The temperature is normalized to the global cluster temperature and the radial coordinate is normalized to the cluster $r_{200}$ radius.}
\label{temp_profs}
\end{figure}
\end{center}

We show in Fig.~\ref{temp_profs} the 3D non-weighted temperature
profile for every cluster in our sample, calculated with the
parameters given in Table \ref{tab_par} and normalized to the
corresponding median temperature from the literature (see
Table~\ref{tab_data}). As for the total mass profiles, the radial
coordinate has been normalized to the corresponding cluster virial
radius. All the profiles are very similar: they slightly increase from
the center, then remain approximately constant to finally decrease at
large radii. The mean temperature profile is consistent with an almost
isothermal distribution for radii smaller than $0.25 r_{200}$ ($\sim$
625 kpc), in agreement with other works (Allen, Schmidt \& Fabian
2001; De Grandi \& Molendi 2002), followed by a rapid decrease for
radii larger than $0.25 r_{200}$, with a logarithmic slope of the
order of $-0.6$ for radii around $0.63\ r_{200}$. This decrease
results to be quite similar to the slope of $-0.64$ found by De Grandi
\& Molendi (2002) at radii larger than $\sim$~750~kpc for their mean
temperature profile, taking into account the cooling and non-cooling
flow clusters of their sample. The results of Markevitch et al. (1998)
based on ASCA observations show similar trends for the temperature
distribution.  The characteristic central drop we obtain in our
temperature profiles is due to the mathematical properties of the
density profile rather than to a real physical effect, but the lack of
resolution of the PSPC impedes from properly addressing this issue,
and we cannot tell anything about the temperature distribution within
the central cooling flow region.

\begin{center}
\begin{figure*}[t]
\centerline{\psfig{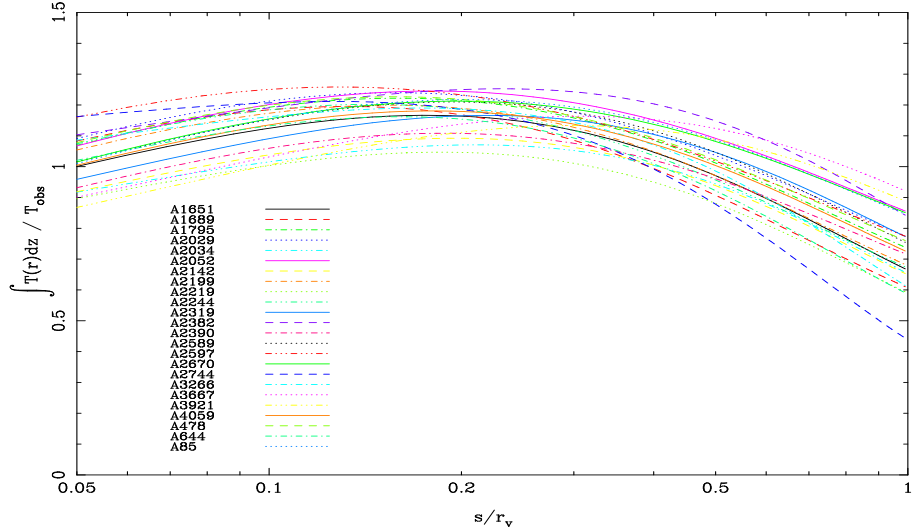}}
\caption{Re-projected emission weighted temperature deduced from our
model. The projected radii $s$ are normalized to the corresponding $r_{v} = r_{200}$ radii, and the temperatures are normalized to the
observed isothermal temperatures.  }
\label{temp_prof_fred}
\end{figure*}
\end{center}

In Fig.~\ref{temp_prof_fred} we show the re-projected emission weighted
temperature profile for every cluster, defined by :
\begin{equation}
     T_{\rm ew}(R) = \frac{\int n^2_{\rm e}(r) T(r) {\rm d}z}
                          {\int n^2_{\rm e}(r) {\rm d}z}
\end{equation}

\begin{figure}
\begin{center}
\psfig{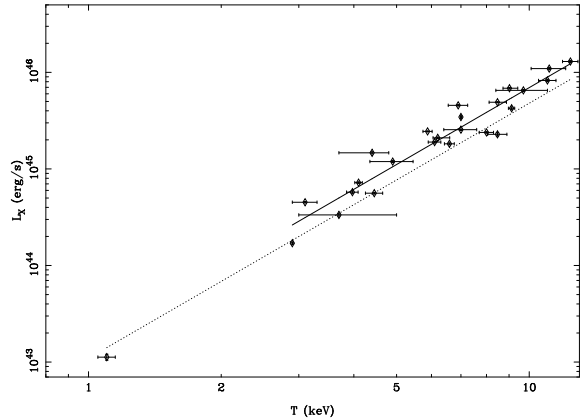}
\caption{Relation between $L_X$ and $T_X$. The point at the lower left
corner corresponds to the group HCG 62. The solid line shows the best
fit (see text) and the dotted line is the fit found by Markevitch 
(1998).}
\label{lx-tx}
\end{center}
\end{figure}

The $L_X-T_X$ relation derived from our calculations is shown in Fig.
\ref{lx-tx}. Our power law fit (excluding HCG 62) is $L_X\propto
T_X^{2.65\pm 0.17}$; the exponent is in agreement with Markevitch
(1998), who finds $L_X\propto T_X^{2.64\pm 0.27}$, but our line is
shifted towards higher luminosities.  This shift is probably due to
the fact that we include the central region in our luminosity
evaluation, while Markevitch excises it.

\subsection{ICM and DM specific entropies}\label{specific_entropy}

\begin{figure}
\begin{center}
\psfig{figure=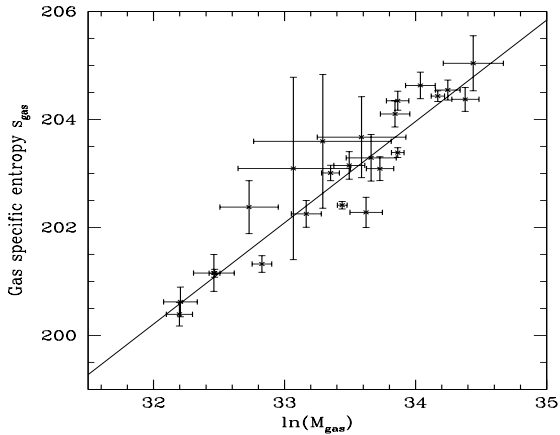,height=6cm,width=8cm}
\caption{Relation between the gas integrated specific entropy,
$s_{gas}$ and the gas mass. The best fit line is indicated (see
text).}
\label{fig_sentropy-mass}
\end{center}
\end{figure}

\begin{figure}
\begin{center}
\psfig{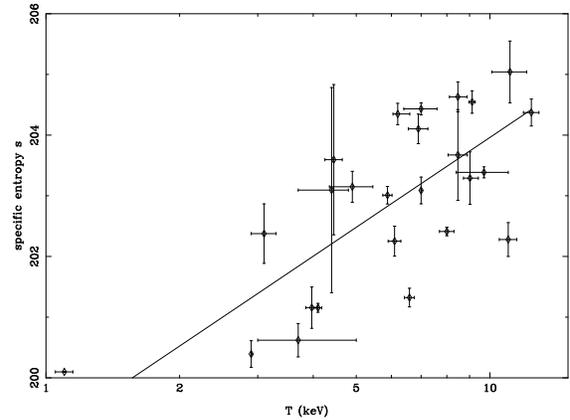}
\caption{Relation between the gas integrated specific entropy
and the mean gas temperature taken from the literature (see text). The
point at the lower left corner corresponds to the group HCG~62.}
\label{fig_sentropy-temp}
\end{center}
\end{figure}

The global specific entropy $s_{gas}$ of the ICM is given in
Table~\ref{tab_phy_par}.  It is found to vary very little from one
cluster to another, as for the specific entropy of stars in elliptical
galaxies (M\'arquez et al. 2001 and references therein). This is,
however, a first order behavior.  Numerical simulations of elliptical
galaxies formed in a hierarchical merging scheme, show that their
specific entropy varies a little with mass, most probably due to
merging processes (Lima Neto et al. 1999, M\'arquez et al. 2000). The
situation is not different in galaxy clusters. We show in
Fig.~\ref{fig_sentropy-mass} that the global specific entropy of
the ICM and the gas mass are indeed clearly correlated, although
this is a second order effect (about 2\%), small compared to the
dominant relation $s_{gas} \sim constant$ that we observe. The
difference with elliptical galaxies is that the slope in
Fig.~\ref{fig_sentropy-mass} is steeper than for ellipticals: if we
write $s_{gas}=s_0+\beta\ln(M_{gas})$, our best fit to the data gives
$\beta =1.86\pm 0.15$ compared to $\beta \simeq 1$ for ellipticals
(with $s_0$ constant). The error bars correspond to $1\sigma$
deviations and were computed from the parameter errors given by MINOS
and by means of Monte Carlo simulations. Every parameter was modeled
by a gaussian distribution with $\sigma$ equal to the corresponding
$1\sigma$ deviation from MINOS. These distributions were then used to
derive the total mass and specific entropy distributions and the
corresponding $1\sigma$ errors for each cluster. For the linear fit we
used a linear least-squares approximation in one dimension, taking
into account the error bars in both directions at the same time.  A
comparable relation is found between the integrated specific entropy
$s_{gas}$ of the ICM and the dynamical mass of the cluster, with
a somewhat steeper slope of $2.67 \pm 0.12$. In our fitting of the
cluster X-ray surface brightness we have also taken into account the
cluster center, thus our estimation of the gas specific entropy
necessarily takes into account the effects of the central cooling
flow. This information is integrated together with the other
non-gravitational thermal processes affecting the intra cluster gas.

The integrated specific entropy of the X-ray gas is displayed in
Fig.~\ref{fig_sentropy-temp} as a function of the observed isothermal
gas temperature. The entropy appears to increase with the temperature,
consistently with a linear increase (the best fit to our data is shown
in Fig.~\ref{fig_sentropy-temp} and corresponds to $s_{gas} \propto
T^{4.92}$). However, the dispersion is too large to assert a linear
dependency. Note that such a relation has already been observed by
Ponman et al. (1999) and Lloyd-Davies et al. (2000), and predicted by
gravitational simulations (Borgani et al. 2001, Muanwong et al. 2002).
The only group shown in this plot does not exhibit a significant
entropy floor, but other groups need to be added.  It would be
important to get a good evaluation of this entropy floor as it is a
strong constraint on the non-gravitational energy injection
(Lloyd-Davies et al. 2000).

The derived gas density and temperature profiles can be used to
compute the gas specific entropy profile for each cluster by
using Eqs.~(\ref{sent_prof}) and (\ref{sent_prof_k}).

\begin{figure}
\begin{center}
\psfig{figure=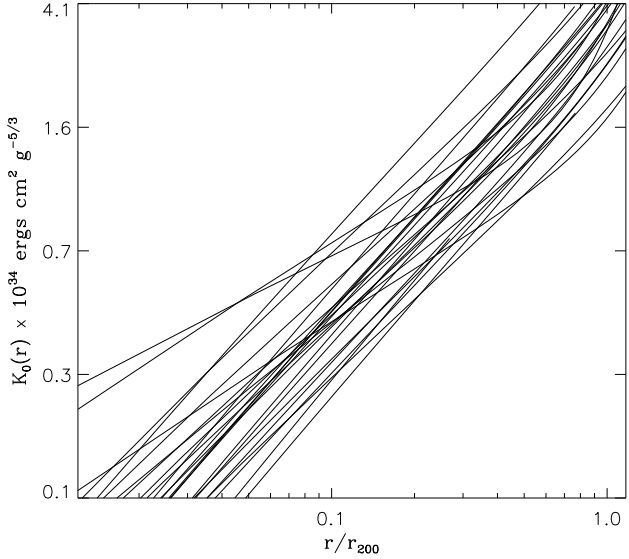,height=8cm}
\caption{$K_0$ profiles based on density and temperature models.}
\label{fig_entropy_prof}
\end{center}
\end{figure}

The $K_0(r)$ profiles are shown in Fig.~\ref{fig_entropy_prof},
where the radial coordinate is normalized to the $r_{200}$ radius. All
these profiles are consistent with a gas specific entropy
increasing with radius, indicating that the gas has been heated by
shock fronts at successive epochs when collapsing into the cluster
potential well, before being virialized. Thus, the specific
entropy profile provides useful information about the ICM thermal
history. Some of the profiles present a clear steepening with radius
at about $0.8\ r_{200}$, while others correspond to power laws
throughout. The average of the profiles defined by
Eq.~(\ref{sent_prof_k}) has a logarithmic slope $d log K_0 / d log r
\sim 1$ at about 0.6 $r_{200}$ and it is in agreement with results
obtained by Tozzi \& Norman (2001) who studied the effect of an
entropy excess in the ICM gas before accretion into the DM halo. No
entropy core is present, perhaps due to the effect of central cooling,
although this could also be due to the mathematical nature of the
profiles derived from the S\'ersic model. In general our profiles are
in agreement with external heating models for rich clusters (Tozzi,
Scharf \& Norman 2000; Tozzi \& Norman 2001). However, we notice that
the $K_0(r)$ profiles obtained from our density and temperature
distributions are about a factor of 7 larger than those inferred by
Tozzi \& Norman (2001). This is probably just due to a
normalization effect related to the way the density and temperature
are computed; it is mainly the entropy variations that are a reliable
physical quantity.

\begin{figure}
\begin{center}
\psfig{figure=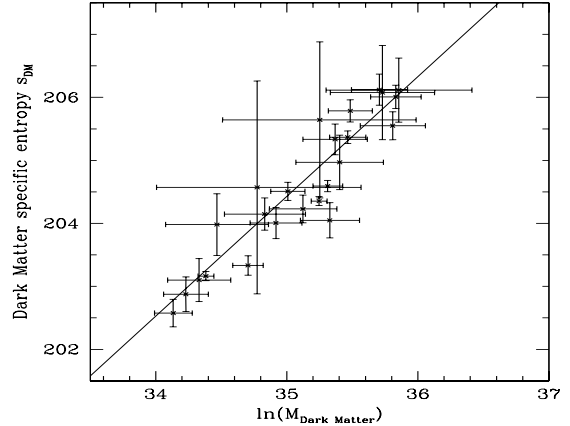,width=8cm,height=6cm}
\caption{Relation between the dark matter integrated specific
entropy and mass.  The best fit line is indicated (see text).}
\label{fig_sdark_mdark}
\end{center}
\end{figure}

The assumed relation (\ref{relation_rapport}) between the dark matter
and gas densities allows us to also compute the global dark matter
entropy (see Table \ref{tab_phy_par}) numerically from
Eq.~(\ref{sent_dm}), which comes out directly from the definition
given in Eq.~(\ref{boltzmann-gibbs-entropy}). The corresponding
entropy--mass relation thus found for the dark matter is very close to
that obtained for the gas, with $\beta \simeq 1.90\pm 0.17$ as seen in
Fig.~\ref{fig_sdark_mdark}.

\begin{center}
\begin{table*}
\caption{Gas and dynamical masses, potential energy, central gas
column density and integrated specific entropy for gas and
DM as derived from S\'ersic parameters.}  \centering
\begin{tabular}{p{1.5cm}*{15}{c}}
\hline
\rule[-2mm]{0mm}{7mm} Cluster & M$_{gas} (\times 10^{14} M_{\odot})$ & $M_{Dyn} (\times 10^{15} M_{\odot})$ & $U_{pot} \times 10^{58}$ (kg m$^2$ s$^{-2})$ & $\Sigma_0$ (kg m$^{-2})$ & Gas Spec. Entr. & DM Spec. Entr.  \\
\hline
\rule[-0mm]{0mm}{4mm}A85   & 5.11$\pm$0.41 & 2.59$\pm$0.35 & 0.78$\pm$0.20 & 0.36$\pm$0.01 & 204.3$\pm$0.18 & 206.1$\pm$0.91 \\
\rule[-0mm]{0mm}{4mm}A478  & 4.98$\pm$0.56 & 2.29$\pm$0.35 & 0.72$\pm$0.21 & 0.68$\pm$0.03 & 204.1$\pm$0.24 & 205.7$\pm$1.00 \\
\rule[-0mm]{0mm}{4mm}A644  & 1.80$\pm$0.13 & 1.18$\pm$0.16 & 0.39$\pm$0.10 & 0.26$\pm$0.01 & 201.3$\pm$0.15 & 203.7$\pm$0.87 \\
\rule[-0mm]{0mm}{4mm}A1651 & 2.53$\pm$0.29 & 1.46$\pm$0.24 & 0.46$\pm$0.14 & 0.27$\pm$0.01 & 202.3$\pm$0.25 & 204.4$\pm$0.98 \\
\rule[-0mm]{0mm}{4mm}A1689 & 4.11$\pm$0.85 & 2.35$\pm$0.52 & 1.05$\pm$0.41 & 0.65$\pm$0.05 & 203.3$\pm$0.43 & 205.3$\pm$1.22 \\
\rule[-0mm]{0mm}{4mm}A1795 & 3.10$\pm$0.20 & 1.63$\pm$0.20 & 0.45$\pm$0.10 & 0.50$\pm$0.01 & 203.0$\pm$0.14 & 204.9$\pm$0.88 \\
\rule[-0mm]{0mm}{4mm}A2029 & 6.14$\pm$0.69 & 3.26$\pm$0.46 & 1.26$\pm$0.31 & 0.66$\pm$0.03 & 204.6$\pm$0.25 & 206.5$\pm$1.00 \\
\rule[-0mm]{0mm}{4mm}A2034 & 4.44$\pm$0.44 & 1.79$\pm$0.42 & 0.64$\pm$0.29 & 0.23$\pm$0.01 & 203.1$\pm$0.22 & 204.6$\pm$0.93 \\
\rule[-0mm]{0mm}{4mm}A2052 & 1.67$\pm$0.38 & 0.95$\pm$0.21 & 0.13$\pm$0.05 & 0.32$\pm$0.02 & 202.4$\pm$0.49 & 204.3$\pm$1.31 \\
\rule[-0mm]{0mm}{4mm}A2142 & 5.11$\pm$0.22 & 2.17$\pm$0.40 & 0.94$\pm$0.35 & 0.38$\pm$0.01 & 203.4$\pm$0.09 & 205.0$\pm$0.80 \\
\rule[-0mm]{0mm}{4mm}A2199 & 1.27$\pm$0.04 & 0.86$\pm$0.08 & 0.18$\pm$0.03 & 0.28$\pm$0.01 & 201.2$\pm$0.07 & 203.5$\pm$0.79 \\
\rule[-0mm]{0mm}{4mm}A2219 & 8.55$\pm$0.85 & 3.56$\pm$0.75 & 2.17$\pm$0.89 & 0.42$\pm$0.01 & 204.4$\pm$0.22 & 205.9$\pm$0.94 \\
\rule[-0mm]{0mm}{4mm}A2244 & 3.92$\pm$1.45 & 3.33$\pm$1.15 & 1.53$\pm$0.88 & 0.36$\pm$0.05 & 203.7$\pm$0.75 & 206.4$\pm$1.60 \\
\rule[-0mm]{0mm}{4mm}A2319 & 7.45$\pm$0.62 & 3.64$\pm$0.63 & 1.79$\pm$0.61 & 0.26$\pm$0.01 & 204.5$\pm$0.19 & 206.4$\pm$0.90 \\
\rule[-0mm]{0mm}{4mm}A2382 & 0.96$\pm$0.01 & 0.67$\pm$0.11 & 0.11$\pm$0.03 & 0.06$\pm$0.01 & 200.4$\pm$0.22 & 202.9$\pm$0.91 \\
\rule[-0mm]{0mm}{4mm}A2390 & 9.06$\pm$2.17 & 3.72$\pm$1.01 & 1.78$\pm$0.85 & 0.62$\pm$0.05 & 205.0$\pm$0.51 & 206.5$\pm$1.31 \\
\rule[-0mm]{0mm}{4mm}A2589 & 0.97$\pm$0.12 & 0.73$\pm$0.11 & 0.13$\pm$0.04 & 0.15$\pm$0.01 & 200.6$\pm$0.28 & 203.2$\pm$1.00 \\
\rule[-0mm]{0mm}{4mm}A2597 & 2.30$\pm$2.14 & 1.26$\pm$1.13 & 0.22$\pm$0.49 & 0.78$\pm$0.52 & 203.1$\pm$1.69 & 204.9$\pm$1.94 \\
\rule[-0mm]{0mm}{4mm}A2670 & 2.86$\pm$2.09 & 2.03$\pm$1.35 & 0.46$\pm$0.59 & 0.22$\pm$0.05 & 203.6$\pm$1.24 & 206.0$\pm$2.18 \\
\rule[-0mm]{0mm}{4mm}A2744 & 4.00$\pm$0.52 & 2.20$\pm$0.47 & 1.36$\pm$0.55 & 0.26$\pm$0.01 & 202.3$\pm$0.28 & 204.4$\pm$0.95 \\
\rule[-0mm]{0mm}{4mm}A3266 & 3.33$\pm$0.11 & 2.03$\pm$0.31 & 0.92$\pm$0.28 & 0.17$\pm$0.01 & 202.4$\pm$0.07 & 204.7$\pm$0.77 \\
\rule[-0mm]{0mm}{4mm}A3667 & 6.93$\pm$0.31 & 2.52$\pm$0.57 & 0.83$\pm$0.38 & 0.19$\pm$0.01 & 204.4$\pm$0.10 & 205.7$\pm$0.81 \\
\rule[-0mm]{0mm}{4mm}A3921 & 3.53$\pm$0.40 & 1.34$\pm$0.31 & 0.29$\pm$0.13 & 0.17$\pm$0.01 & 203.1$\pm$0.25 & 204.5$\pm$0.98 \\
\rule[-0mm]{0mm}{4mm}A4059 & 1.24$\pm$0.20 & 0.80$\pm$0.14 & 0.15$\pm$0.05 & 0.22$\pm$0.01 & 201.2$\pm$0.34 & 203.5$\pm$1.09 \\
\hline
\end{tabular}
\label{tab_phy_par}
\end{table*}
\end{center}

\subsection{Potential energy -- mass relation}\label{scaling_relation}

\begin{figure}
\begin{center}
\psfig{figure=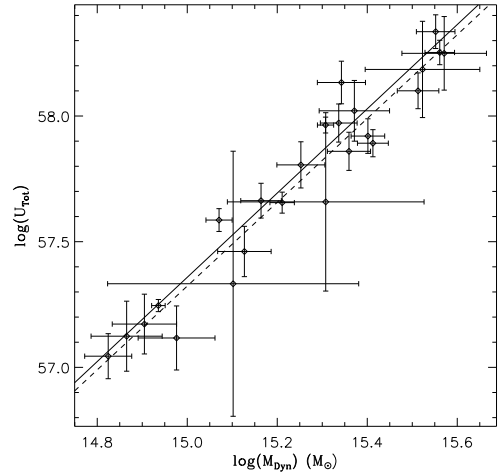,height=7cm}
\caption{Relation between the dynamical mass and the potential energy.
The solid line corresponds to the best fit to the data and the dashed
line is the theoretical slope of $I=5/3$ defined by the relation
$ln(U_{pot})-I\ ln(M_{dyn})=const$.
\label{fig_mdyn_epot}}
\end{center}
\end{figure}

The relation between the dynamical mass and potential energy,
previously shown for elliptical galaxies (M\'arquez et al. 2001) and
in numerical simulations (Lanzoni, 2000; Jang-Condell \& Hernquist,
2001), is displayed in Fig.~\ref{fig_mdyn_epot} for our set of
clusters.  If we write it as $ln (U_{pot}) - I\,ln (M_{dyn}) = const$,
the best fit to our data imply $I =1.68 \pm 0.08$, in excellent
agreement with the theoretical value of $5/3$. The error bars and
linear fit are computed in the same way as explained in the previous
section.

\subsection{Discussion}
\label{discussion}

The correlations presented in the previous sections between the mass,
potential energy and integrated specific entropy, confirm the existence of
an entropic surface, $s(\Sigma_0,a,\nu) - \beta \ ln (M
(\Sigma_0,a,\nu))= const$ (see Eq. (\ref{boltzmann-gibbs-entropy})),
and a potential energy-mass surface, $ln (U_{pot}(\Sigma_0,a,\nu)) - I
\ ln (M_{dyn}(\Sigma_0,a,\nu)) = const$, in the S\'ersic parameters
space, remarkably similar to the case of elliptical galaxies
(M\'arquez et al. 2001). Indeed, we observe the clusters in our sample
to be located at the intersection of these two surfaces, indicating
the existence of an ``entropic line'' for galaxy clusters (see Magnard
2002).
We have checked by simulations that this line is also recovered for a
random set of values of the S\'ersic parameters, suggesting the
possibility that this relation could be a consequence of the model
assumed to describe the final viralized system (in our case a S\'ersic
profile). However, the fact that a S\'ersic profile reproduces very
well the X-ray surface brightness of clusters, and hence the gas
distribution of the ICM, supports the idea that the physical processes
operating during the formation and evolution of galaxy clusters, which
are of course responsible for the final structure reached by the ICM
and DM halo, are indeed at the origin of the entropic line. To confirm
this, the same method as in this work, but with different models for
the gas density profile should be used.  Moreover, the existence of
both an entropic surface and a potential energy-mass surface for
galaxy clusters implies that these objects can be considered as a
single-parameter family, described by one of the S\'ersic parameters
only (e.g. M\'arquez et al. 2001). Interestingly, analogous
correlations have been obtained for galaxy clusters by Fujita \&
Takahara (1999).
The importance of the above mentioned correlations resides in the fact
that they are probably the result of the physics ruling cluster
formation. A correlation between the global specific entropy and the
mass conserves information on the various events affecting the
thermodynamical history of clusters. The observed variation of
$s_{gas}$ with dynamical mass in clusters suggests that dissipating
processes in clusters play an important role as generators of
entropy. These mainly correspond to Bremsstrahlung emission ($L
\propto M^{4/3}$) and cooling flows. Merging processes between
clusters are of importance in such a relation, because of their impact
on the final total mass and on the amount of entropy produced during
the cluster formation. Violent merger events can be accompanied by an
important dissipation of energy and creation of entropy, while minor
mergers can be translated in an adiabatic accretion of matter without
a significant production of entropy. These energy losses, however, are
all negligible compared to the cluster gravitational
energy. Thus the value of the slope $\beta$ in the specific
entropy-mass relation reflects the impact of such processes on the
cluster history. Clusters with higher global specific entropy could
have undergone more episodes of hierarchical merging through their
histories, thus becoming more massive.

On the other hand, considering the collapse of matter to form a
virialized gravitational system, the correlation between the potential
energy and the total mass of the final structure is a natural
consequence of the conservation of energy and mass during its
formation. A self-similar relation defined by $U \propto M^{5/3}$ is
expected from theory (see M\'arquez et al. 2001) and we show that it
is indeed also followed by our observed galaxy clusters.

All these results strongly suggest that the formation processes
affecting galaxies and clusters of galaxies are quite similar
regardless the scale involved.  

\section{Conclusions}\label{conclusions}

We have shown in the present work that the S\'ersic profile can be
used as a good tracer of the matter distribution in clusters, under
the assumption that clusters are well relaxed structures, as it was
the case in elliptical galaxies. Its mathematical properties make it
an interesting and useful tool that can be employed to explore the
physics of relaxed systems, although any other appropriate
profile can be used. The density profiles obtained here reproduce
well the X-ray surface brightness profiles of the ROSAT PSPC
images. The asymptotic behaviour of these profiles towards the cluster
center turns out to be shallower than the NFW profile, but still
within the limits predicted by numerical simulations concerning the
central slope of galaxy-sized DM halos. Temperature profiles derived
here (considering the hydrostatical equilibrium hypothesis for the
cluster structure) are in agreement with other works. We estimate the
integrated specific entropy content for galaxy clusters and our specific
 entropy profiles are consistent with the predicted shape of
the entropy distribution for massive clusters, obtained by simulations
which take into account pre-heating and cooling processes.

We have shown that both for the gas and for the dark matter the integrated
 specific entropy and the potential energy have behaviors
comparable to those observed for stars in elliptical galaxies: the
integrated specific entropy is constant to first order and in reality
increases slightly with mass as a logarithmic function, while the
potential energy scales with mass as a power law. Note that the index
of this power law is close to the theoretical value of 5/3 for
elliptical galaxies and clusters. This strongly suggests that all
these self gravitating systems behave similarly, even though they may
have very different masses and thermodynamical histories. Elliptical
galaxies could then be considered as scaled down versions of galaxy
clusters (Moore et al.  1999). Moreover, integrated specific
entropy-mass and potential energy-mass correlations should be the
result of the formation history of the clusters. Heating mechanisms
and merger events play an important role here and total mass and
energy are conserved during the whole formation process of the final
virialized structure.

It would be interesting to apply the S\'ersic model to high redshift
galaxy clusters in order to test the possible evolution of the scaling
relations found in this work. The use of Chandra and XMM-Newton data
will be crucial in these kinds of studies due to their higher
resolution and sensitivity compared to ROSAT. We also note that a
similar analysis could be carried out in samples of synthetic clusters
derived from numerical simulations.

\begin{acknowledgements}

We are very grateful to Daniel Gerbal, Gast\~ao B.  Lima Neto, Gary
Mamon and Sergio Dos Santos for many enlightening discussions.
R.D. acknowledges many interesting discussions with Rapha\"el Lescouz\`eres.

\end{acknowledgements}


\begin{thebibliography}{}

\bibitem {} Allen S., Ettori S. \& Fabian A. 2001, MNRAS 324, 877
\bibitem {} Allen S., Schmidt R. \& Fabian A. 2001, MNRAS 328, L37
\bibitem {} Arnaud M., Aghanim N., Gastaud R. et al. 2001, A\&A 365, L67 
\bibitem {} Arnaud M. \& Evrard A. 1999, MNRAS 305, 631
\bibitem {} Balogh M. L., Babul A. \& Patton D. R. 1999, MNRAS 307, 463
\bibitem {} Bertschinger E. 1985, ApJS, 58, 39
\bibitem {} Binney J. \& Evans N. 2001, MNRAS 327, L27
\bibitem {} Binney J. \& Tremaine S. 1987, ``Galactic Dynamics'', Princeton
University Press
\bibitem {} Bonnor W. B. 1956, MNRAS, 116, 351 
\bibitem {}  {Borgani} S., {Governato} F., {Wadsley} J. et al. 2001, ApJL 559, L71
\bibitem {} Bower R.G., Benson A.J., Lacey C.G. et al. 2001, MNRAS 325, 497
\bibitem {} Brighenti F. \& Mathews W. 2001, ApJ 553, 103
\bibitem {} Bryan G. 2000, ApJ 544, L1
\bibitem {} Caon N., Capaccioli M. \& D'Onofrio M. 1993, MNRAS 265, 1013
\bibitem {} Cavaliere A. \& Fusco-Femiano R. 1976, A\&A 49, 137
\bibitem {} Ciotti L. \& Bertin G. 1999, A\&A 352, 447
\bibitem {} de Vaucouleurs G. 1948, Ann. d'Astroph. 11, 247
\bibitem {} De Grandi S. \& Molendi S. 2002, ApJ 567, 163
\bibitem {} Dos Santos S. \& Dor\'e O. 2002, A\&A 383, 450
\bibitem {} Durret F., Gerbal D., Lachi\`eze-Rey M., Lima Neto G.B. \&
Sadat R. 1994, A\&A 287, 733
\bibitem {} Ebeling H., Voges W., B\"ohringer H. et al. 1996, MNRAS 281, 799
\bibitem {} Ettori S., De Grandi S. \& Molendi S. 2002, A\&A 391, 841
\bibitem {} Fillmore J. \& Goldreich P. 1984, ApJ 281, 1
\bibitem {} Flores R. \& Primack J. R. 1994, ApJ 427, L1
\bibitem {} Frenk C. S., White S.D.M., Bode P. et al. 1999, ApJ 525, 554 
\bibitem {} Fujita Y. \& Takahara F. 1999, ApJ 519, L51
\bibitem {} Gerbal D., Durret F., Lima Neto G.B., Lachi\`eze-Rey M.
1992, A\&A 253, 77
\bibitem {} Gerbal D., Lima Neto G. B., M\'arquez I., Verhagen H. 1997,
MNRAS 285, L41
\bibitem {} Gunn J. \& Gott J. R. III 1972, ApJ 176, 1
\bibitem {} Helsdon S.F. \& Ponman T.J. 2000, MNRAS 315, 356
\bibitem {} Hicks A., Wise M., Houck J. \& Canizares C. 2002,
ApJ 580, 763
\bibitem {} Irwin J.A., Bregman J.N. \& Evrard A. E. 1999, ApJ 519, 518 
\bibitem {} Irwin J.A. \& Bregman J.N. 2000, ApJ 538, 543
\bibitem {} James F. 1994, CERN Program Library Long Writeup D506
\bibitem {} Jang-Condell H. \& Hernquist L. 2001, ApJ 548, 68
\bibitem {} Jing Y. P. \& Suto Y. 2000, ApJ 529, L69
\bibitem {} Lanzoni B., 2000, PhD Thesis, Universit\'e Paris 7
\bibitem {} Lima Neto G. B., Gerbal D., M\'arquez I. 1999, MNRAS 309, 481
\bibitem {} Lloyd-Davies E.J., Ponman T.J. \& Cannon D.B. 2000, MNRAS 315, 689
\bibitem {} Loken C., Norman M.L., Nelson E. et al. 2002, ApJ 579, 571
\bibitem {} Magnard F. 2002, PhD Thesis, Universit\'e Paris 6
\bibitem {} Markevitch M., Forman W. R., Sarazin C. L. \& Vikhlinin A. 1998,
ApJ 503, 77
\bibitem {} Markevitch M. 1998, ApJ 504, 27
\bibitem {} M\'arquez I., Lima Neto G.B., Capelato H., Durret F.,
Gerbal D. 2000, A\&A 353, 823
\bibitem {} M\'arquez I., Lima Neto G.B., Capelato H. et al. 2001,
A\&A 379, 767
\bibitem {} Mellier Y. \& Mathez G. 1987, A\&A 175, 1
\bibitem {} Mohr J. J., Mathiesen B., Evrard A. E. 1999, ApJ 517, 627
\bibitem {} Moore B., Quinn T., Governato F., Stadel J. \& Lake G. 1999,
MNRAS 310, 1147
\bibitem {} {Muanwong} O., {Thomas} P.~A., {Kay} S.~T. \& {Pearce} F.~R. 2002, MNRAS 336, 527
\bibitem {} Navarro J., Frenk C. \& White S. 1996, ApJ 462, 563
\bibitem {} Navarro J., Frenk C. \& White S. 1997, ApJ 490, 493
\bibitem {} Ponman T. J., Cannon D. B. \& Navarro J. F. 1999, Nature, 397, 135
\bibitem {} Renzini A. 1997, ApJ 488, 35
\bibitem {} Rosati P., Borgani S., Norman C. 2002, ARA\&A 40, 539
\bibitem {} Sarazin C. 1988, ``X-Ray Emission from Clusters of Galaxies'',
Cambridge University Press
\bibitem {} Schindler S. 1999, A\&A 349, 435
\bibitem {} S\'ersic J. L. 1968, Atlas de Galaxias Australes, Observatorio
Astron\'omico de C\'ordoba, Argentina
\bibitem {} Snowden S. L., McCammon D., Burrows D.N., Mendenhall J.A. 1994, 
ApJ 424, 714
\bibitem {} Subramanian K. 2000, ApJ 538, 517
\bibitem {} Tamura N., Kaastra J.S., Peterson J.R. et al. 2001, A\&A 365, L87 
\bibitem {} Taylor J. E. \& Navarro J. F. 2001, ApJ 563, 483
\bibitem {} Teyssier R., Chi\`eze J.-P. \& Alimi J.-M. 1997, ApJ 480, 36
\bibitem {} Teyssier R. 2002, A\&A 385, 337
\bibitem {} Tozzi P., Scharf C. \& Norman C. 2000, ApJ 542, 106
\bibitem {} Tozzi P. \& Norman C. 2001, ApJ 546, 63
\bibitem {} van den Bosch F. C. \& Swaters R. A. 2001, MNRAS 325, 1017 
\bibitem {} White D. 2000, MNRAS 312, 663
\bibitem {} Wu X.-P., Xue Y.-J, Fang L.-Z. 1999, ApJ 524, 22


\end{thebibliography}
\end{document}